\documentclass[onecolumn,showpacs]{article}
\usepackage{graphicx}
\usepackage{dcolumn}
\usepackage{bm}
\usepackage{amsmath}
\usepackage{amssymb}
\usepackage{multirow}
\usepackage{caption}
\usepackage{subcaption}
\usepackage{epstopdf}
\usepackage{hyperref}
\begin{document}
{\setlength{\oddsidemargin}{1.2in}
\setlength{\evensidemargin}{1.2in} } \baselineskip 0.55cm
\begin{center}
{\LARGE {\bf Stability Analysis of Cosmological models in $f(T,\phi)$ Gravity}}
\end{center}
\date{\today}
\begin{center}
  Amit Samaddar, S. Surendra Singh \\
   Department of Mathematics, National Institute of Technology Manipur,\\ Imphal-795004,India\\
   Email:{ samaddaramit4@gmail.com, ssuren.mu@gmail.com }\\
 \end{center}

 \textbf{Abstract} We investigated the stability condition in $f(T,\phi)$ gravity theory for considering two models by using dynamical system. We assume the forms of $G(T)$ are $(i)$ $G(T)$ = $\alpha T+\frac{\beta}{T}$, $(ii)$ $G(T)$ = $\zeta T$ ln$(\psi T)$, where $\alpha$, $\beta$, $\zeta$ and $\psi$ be the free parameters. We evaluated the equilibrium points for these models and examine the stability behavior. We found five stable critical points for Model I and three stable critical points for Model II. The phase plots for these systems are examined and discussed the physical interpretation. We illustrate all the cosmological parameters such as $\Omega_{m}$, $\Omega_{\phi}$, $q$ and $\omega_{Tot}$ at each fixed points and compare the parameters with observational values. Further, we assume  hybrid scale factor and the equation of redshift and time is $t(z)=\frac{\delta}{\sigma}W\bigg[\frac{\sigma}{\delta}\bigg(\frac{1}{a_{1}(1+z)}\bigg)^{\frac{1}{\delta}}\bigg]$. We transform all the parameters in redshift by using this equation and examine the behavior of these parameters. Our models represent the accelerating stage of the Universe. The energy conditions are examined in terms of redshift and SEC is not satisfied for the model. We also find the statefinder parameters $\{r,s\}$ in terms of z and discuss the nature of $r-s$ and $r-q$ plane. For both pairs $\{r,s\}$ and $\{r,q\}$ our models represent the $\Lambda$CDM model. Hence, we determine that our $f(T,\phi)$ models are stable and it satisfies all the observational values. \\


  \section{  Introduction}\label{sec1}

 From current observation of Supernova Type Ia (SNeIa) \cite{Riess09}, Cosmic Microwave Background (CMB)\cite{Perlmutter99}, Baryon Acoustic Oscillations (BAO)\cite{Spergel03}, it is provided that accelerating stage of the Universe is undergoing. The matter existing in the Universe are affected by an exotic forms of energy which is known as dark energy (DE) \cite{Cole05}. From latest CMBR data, our Universe contains $76\%$ dark energy. In the right hand side of Einstein's field equations, the cosmological constant has negative equation of state ($\omega_{de}$) which is accountable for the late time acceleration of the Universe. The cosmological constant creates several problems like cosmological constant problem. In present days, the combination of energy density and cosmological constant is known as critical density and the value is $\rho_{\Lambda}\sim10^{-47}GeV^{4}$ but in quantum theory, the energy density is $10^{121}$ times bigger than the observational value and the value is $\rho_{\Lambda}\sim10^{74}GeV^{4}$ \cite{Sami06}. The form of the equation of state (EoS) parameter is $\omega_{de}=\frac{p_{de}}{\rho_{de}}$ \cite{Kowalski08,Spergel07}. If the EoS parameter $\omega_{de}$ tends to $-1$, then it represents the standard cosmology. If $\omega_{de}=1$ then it describes the stiff fluid, if $\omega_{de}=0$ then it describes matter dominant phase while $\omega_{de}=\frac{1}{3}$ describes radiation dominant phase, for $-1<\omega\leq -\frac{1}{3}$ the Universe is in quintessence phase and universe also exhibits for phantom dark energy model $\omega_{de}<-1$ and lastly $\omega_{de}=-1$ represents the cosmological constant $i.e.$ $\Lambda$CDM model \cite{Komatsu11}. To discuss the expansion behavior of the Universe, cosmologists proposed a new parameter which is called deceleration parameter $(q)$. If the numerical value of deceleration parameter $q>0$, then it represents the decelerated epoch of the Universe while for $q<0$, it represents the accelerated epoch of the Universe but for $q=0$, it represents the marginal expansion of the Universe. To solve the cosmological constant problem, we have an alternate theory which is dynamical dark energy section with insertion of the scalar field for example quintessence \cite{Wetterich88,Ratra88}, k-essence \cite{Chiba00,Mukhanov00}, Galileons \cite{Nicolis09,Deffayet09} etc.\\

 To understand the several perspectives of modern cosmology, cosmologists need the modification of General Relativity. Several methods are introduced to explain the observational evidences. The easiest way to modify the Einstein's gravity theory is to substitute the Ricci scalar with a function $f(R)$ in Einstein-Hilbert action \cite{Capozziello03}. An identical formulation of General Relativity is teleparallel gravity where instead of curvature by torsion which is responsible for the gravitational interaction \cite{Mirza17}. Einstein first introduced this model to establish the electromagnetism and the Weitzenb$\ddot{o}$ck connection of non-Riemannian manifold in gravity theory \cite{Einstein28}. Teleparallel Gravity (TG) can be assumed as a gauge theory \cite{Andrade00} but General Relativity is related to geometric phenomenon. However, the General Relativity and Teleparallel Gravity both are not same but the equations of motion are same thus the new term Teleparallel equivalent of General Relativity (TEGR) is introduced. The alternation of Teleparallel equivalent of General Relativity is $f(T)$ gravity where the torsion scalar $T$ is replaced by an arbitrary function $f(T)$ which is same as $f(R)$ gravity theory \cite{Bengochea09}. The field equations of modified teleparallel gravity are second order but $f(R)$ gravity has fourth order field equations but Local Lorentz Invariance does not exist here. $f(T)$ gravity successfully describes the acceleration stage of the Universe, so, in present days, this is the most interesting theory. The another interesting theory is developed by adding scalar field with torsion which is called scalar torsion theory or $f(T,\phi)$ gravity. Several authors already analyzed the various perspectives of $f(T,\phi)$ gravity in different ways which are Noether symmetry of $f(T,\phi)$ gravity discussed by \cite{Levi22}. Cosmological dynamics of dark energy of $f(T,\phi)$ gravity analyzed by \cite{Manuel21}. Dynamical system in  $f(T,\phi)$ gravity discussed by \cite{Kadam23} etc.\\

 Dynamical system is very useful technique in cosmology to solve the Einstein's field equations. In this manuscript, we discussed the stability analysis in $f(T,\phi)$ gravity \cite{Amit23}. One of the major issues in theories of gravity is to find the analytical or numerical solutions due to the complicated field equations \cite{Singh19}. Some nonlinear terms are present in Einstein's Field equations which are not easy to solve and hence comparison to the observations cannot be easy one. To solve the Einstein's equations, some other methods are required and dynamical system analysis is one of such method which is capable to solve the nonlinear terms in Einstein's equations. Dynamical system is used to find the numerical solutions and understand the stability behavior of a given system \cite{Singh23}. In dynamical system, the most important thing is to find the critical points from the set of autonomous first-order ordinary differential equations which are obtained from the Einstein's equations. The stability analysis of a model are evaluated by calculating the Jacobian matrix at each equilibrium points and finding the characteristic values from Jacobian matrix \cite{Shah21}. This is the process to analyze the stability behavior of any model near a critical point \cite{Sonia22}.\\

 The outlines of this manuscript are: the field equations of $f(T,\phi)$ gravity are presented in $f(T,\phi)$ gravity in sec. (\ref{sec2}). In sec. (\ref{sec3}), we discuss the stability behavior of two model by introducing some new variables from the field equations. In this section, we find the equilibrium points and explore the stability behavior by phase plots. In sec. \ref{sec4}, we explore the nature of some physical parameters like energy density, pressure, deceleration parameter with the help of hybrid expansion law. In sec. (\ref{sec5},\ref{sec6}), we discussed the energy conditions and the nature of the statefinder parameter. Conclusions are given in sec. (\ref{sec7}).\\

 \section{$f(T,\phi)$ Gravity and field equations}\label{sec2}

The substitute representation of gravity with respect to the torsion scalar but not curvature is Teleparallel Gravity \cite{Arcos04,Pereira12}. In modified teleparallel gravity, the geometric part of the action is an algebraic function that depends on the torsion scalar $(T)$ which is known as $f(T)$ gravity. In TEGR action, the torsion scalar is promoted by an arbitrary function with the addition of the scalar field $\phi$ which is the generalization of $f(T)$ gravity. The action of the $f(T,\phi)$ gravity with the presence of matter and radiation is \cite{Otalora20}
 \begin{equation}\label{1}
   S=\; \int d^{4}xe[f(T,\phi)+P(\phi)X]+S_{m}+S_{r},
 \end{equation}
  where $f(T,\phi)$ describes the function of torsion scalar $(T)$ and scalar field $(\phi)$, $e=det(e^{C}_{\mu})=\sqrt{-g}$ and $X=-\partial^{\mu}\phi \partial_{\mu}\phi/2$ is the kinetic term multiplication by an arbitrary function $P(\phi)$ in the action. General Relativity can be revealed in the structure of teleparallel gravity by applying the tetrad $(e^{C}_{\mu})$ and spin connection $(\omega^{C}_{D\mu})$ instead of the metric tensor. The observable component is tetrad field $e^{C}_{\mu}$ and the relation between Minkowski tangent space metric $\eta_{CD}$ and metric tensor $g_{\mu\nu}$ is $g_{\mu\nu}=\eta_{CD}e^{C}_{\mu}e^{D}_{\nu}$, where $\eta_{CD}$ $=$ $diag (-1,1,1,1)$. The tetrad component fulfills the orthogonality condition $e^{\mu}_{C}e^{D}_{\mu}=\delta^{D}_{C}$. The torsion scalar can be defined as,
  \begin{equation}\label{2}
    T= S^{\mu\nu}_{\psi}T^{\psi}_{\mu\nu},
  \end{equation}
  where $T^{\psi}_{\mu\nu}$ and $S^{\mu\nu}_{\psi}$ illustrate the tensor of the torsion and superpotential. The superpotential tensor is defined by,
  \begin{equation}\label{3}
    S^{\mu\nu}_{\psi}= \frac{1}{2}(K^{\mu\nu}_{\psi}+\delta^{\mu}_{\psi}T^{\beta\nu}_{\beta}-\delta^{\nu}_{\psi}T^{\beta\mu}_{\beta}),
  \end{equation}
 where the contortion tensor $K^{\mu\nu}_{\psi}=\frac{1}{2}(T^{\mu\nu}_{\psi}+T^{\nu\mu}_{\psi}-T^{\nu\mu}_{\psi})$. In equation \ref{2}, the torsion tensor $T^{\psi}_{\mu\nu}$ is defined by,
 \begin{equation}\label{4}
  T^{\psi}_{\mu\nu}=e^{\psi}_{C}\partial_{\mu}e^{C}_{\nu}-e^{\psi}_{C}\partial_{\nu}e^{C}_{\mu}+e^{\psi}_{C}\omega^{C}_{D\mu}e^{D}_{\nu}-e^{\psi}_{C}\omega^{C}_{D\nu}e^{D}_{\mu}.
 \end{equation}
  It is attached with the Weitzenb$\ddot{o}$ck connection of the teleparallel gravity. The field equations are derived by varying the tetrad $e^{C}_{\mu}$ in the action or by using the relation of the torsion scalar $(T)$ and curvature $(R)$ with Levi-Civita connection and contortion tensor which satisfy
  \begin{equation}\label{5}
    T=-R+e^{-1}\partial\mu(eT^{\beta\mu}_{\beta}).
  \end{equation}
  and hence, the field equations of General Relativity and Teleparallel Gravity are same. We assume the flat, isotropic and homogeneous Friedmann-Lema$\hat{i}$tre-Robertson-Walker metric to derive the $f(T,\phi)$ gravity field equations as,
  \begin{equation}\label{6}
    ds^{2}= -dt^{2}+a^{2}(t)(dx^{2}+dy^{2}+dz^{2}),
  \end{equation}
  where $a(t)$ is the scale factor and we assume the diagonally tetrad field $e^{C}_{\mu}=diag(-1,a,a,a)$. The Friedmann equations and Klein-Gordon equation of $f(T,\phi)$ gravity can be derived by varying the action of equation (\ref{1}) with tetrad field and scalar field as follows,
  \begin{equation}\label{7}
    f(T,\phi)-P(\phi)X-2Tf_{,T}=\rho_{r}+\rho_{m},
  \end{equation}
  \begin{equation}\label{8}
  f(T,\phi)+P(\phi)X-2Tf_{,T}-4\dot{H}f_{,T}-4H\dot{f}_{,T}=-p_{r},
  \end{equation}
  \begin{equation}\label{9}
    -P_{,\phi}X-3P(\phi)H\dot{\phi}-P(\phi)\ddot{\phi}+f_{,\phi}=0
  \end{equation}
 where $H=\frac{\dot{a}}{a}$ is the Hubble parameter and $\dot{a}$ represents the derivative with respect to time $t$, $f_{,T}=\frac{\partial f}{\partial T}$ and $f_{,\phi}=\frac{\partial f}{\partial \phi}$. $\rho_{m}$ and $\rho  _{r}$ represents the matter and radiation energy density and $p_{r}$ is the pressure for radiation. We use the expression of torsion scalar in terms of Hubble parameter is $T=6H^{2}$. We assume the form of the function $f(T,\phi)$ as \cite{Hohmann18},
 \begin{equation}\label{10}
   f(T,\phi)=-\frac{T}{2k^{2}}+G(T)-V(\phi),
 \end{equation}
 where $G(T)$ is the function of $T$ and $V(\phi)$ is the scalar potential. The equation of state parameter for matter dominant universe is $\omega_{m}=\frac{p_{m}}{\rho_{m}}=0$, for radiation $\omega_{r}=\frac{p_{r}}{\rho_{r}}=\frac{1}{3}$ epoch. Now equations (\ref{7}-\ref{9}) can be written as,
 \begin{equation}\label{11}
   \frac{3}{k^{2}}H^{2}= 2TG_{,T}-G(T)+V(\phi)+P(\phi)X+\rho_{m}+\rho_{r},
 \end{equation}
 \begin{equation}\label{12}
   -\frac{2}{k^{2}}\dot{H}=-4\dot{H}(G_{T}+2TG_{,TT})+2P(\phi)+\rho_{m}+\frac{4}{3}\rho_{r},
 \end{equation}
 \begin{equation}\label{13}
   P_{,\phi}(\phi)X+P(\phi)\ddot{\phi}+3P(\phi)H\dot{\phi}+V_{,\phi}(\phi)=0.
 \end{equation}
 By \cite{M06}, the Friedmann equations (\ref{11}) and (\ref{12}) can be written as,
 \begin{equation}\label{14}
   \frac{3}{k^{2}}H^{2}=\rho_{r}+\rho_{m}+\rho_{de},
 \end{equation}
 \begin{equation}\label{15}
  -\frac{2}{k^{2}}\dot{H}=p_{de}+\rho_{de}+\rho_{m}+\frac{4}{3}\rho_{r}.
 \end{equation}
 To compare the equations (\ref{14}) and (\ref{15}) with the equations (\ref{11}-\ref{12}), the expressions of pressure and energy density of the dark energy can be obtained as,
 \begin{equation}\label{16}
   \rho_{de}= 2TG_{,T}-G(T)+V(\phi)+P(\phi)X,
 \end{equation}
 \begin{equation}\label{17}
   p_{de}= -2TG_{,T}+G(T)-4\dot{H}(G_{T}+2TG_{,TT})-V(\phi)+P(\phi)X,
 \end{equation}
 We consider potential energy $V(\phi)=V_{0}e^{-\lambda\phi}$ and $P(\phi)=1$. To perform the dynamical system analysis from the above two equations (\ref{16}) and (\ref{17}), we need to assume some special form of $G(T)$ and in this work, we assume two $G(T)$ forms which are given in \ref{sec3}. The energy equations are as follows:
 \begin{equation}\label{18}
   \dot{\rho_{m}}\;+3H\rho_{m}=0,
 \end{equation}
 \begin{equation}\label{19}
   \dot{\rho_{r}}\;+4H\rho_{r}=0.
 \end{equation}
 The continuity equation for the pressure and energy density of the dark energy is
 \begin{equation}\label{20}
   \dot{\rho_{de}}+3H(\rho_{de}+p_{de})=0.
 \end{equation}
 \section{Stability Analysis of $f(T,\phi)$ gravity models}\label{sec3}
 The main objective to learn the dynamical system, specially for non-linear equations is to visualize the stability conditions of the fixed points or equilibrium points. Dynamical system is the most necessary technique to learn cosmological behavior in the Universe, where we could not find the exact solutions due to the complicated systems \cite{Rafael17}. The dynamical systems are mostly used in cosmological models for non-linear systems of differential equations. The form of the system is $\dot{v}$ $=$ $\sigma(v)$, where the function $\sigma:V\rightarrow V$, $\dot{v}$ be the derivative with respect to time $t \in \mathbb{R}$ and $v=(v_{1},v_{2},v_{3},....,v_{m}) \in V$, $\sigma(v)=(\sigma_{1}(v),\sigma_{2}(v),....,\sigma_{m}(v))$ \cite{Wright29}. This shows that we analyze the stability conditions for $m$ variables with $m$ equations. The equation $\dot{v}$ $=$ $\sigma(v)$ said that the rate of change $\frac{dv}{dt}$ for function $v(t)$ with some condition. The condition is: if the current value is $v$, then the rate of change is $\sigma(v)$. The equation $\dot{v}$ $=$ $\sigma(v)$ is called ordinary differential equation. The differential equation is called autonomous if the condition doesn't depend upon time $t$, it only depends about the current value of the variable $v$. $v=v_{0}$ be the fixed point of the system $\dot{v}$ $=$ $\sigma(v)$ if and only if $\sigma(v_{0})=0$ \cite{Wright30}. We analyze the stability behavior of fixed points. The fixed point $v_{0}$ is stable if $\forall$ $\eta>0$ $\exists$ a $\xi$ such that $\phi(t)$ be the solution of $\dot{v} = \sigma(v)$ that satisfies the condition $\| \phi(t_{0})-v_{0}\|<\xi$, then $\phi(t)$ exist $\forall$ $t\geq t_{0}$ and satisfy $\| \phi(t)-v_{0}\|<\eta$ $\forall$ $t\geq t_{0}$. The critical point $v_{0}$ is asymptotically stable if $\exists$ a $\xi$ s.t. $\phi(t)$ be the solution of $\dot{v}$ $=$ $\sigma(v)$ satisfying $\|\phi(t_{0})-v_{0}\|<\xi$ then $\lim_{t\rightarrow\infty}\phi(t)$ $=$ $v_{0}$ \cite{Biswas15}. The minimal distinction between stable and asymptotic stable critical point is that for asymptotic critical point all trajectories approach to the point, but for stable critical point all trajectories made a circle near at that point \cite{Chaubey32}. In cosmology, all stable critical points are treated as asymptotically stable fixed point. The critical points which are not stable is called unstable critical points i.e. the trajectories starting near the critical points and escape away from it. Now we introduce some approaches to understand  the stability criteria at the fixed points. Linear stability theory is the most useful process to analyze the physical properties of cosmological models \cite{Campo33}. This theory is used to linearize the equations at the equilibrium point for studying the dynamical properties near this point. Assume that $v_{0}$ is the equilibrium point of the system $\dot{v}$ $=$ $\sigma(v)$. At equilibrium points, the system $\dot{v}$ =$ \sigma(v)$ linearized by Taylor's expansion where each component of the vector field $\sigma(v)=(\sigma_{1}(v),\sigma_{2}(v),....,\sigma_{m}(v))$, becomes such that
 \begin{equation}\label{21}
   \sigma_{i}(v)\; =\; \sigma_{i}(v_{0})+\sum^{n}_{j=1}\frac{\partial \sigma_{i}}{\partial v_{j}}(v_{0})y_{j}+\frac{1}{2!} \sum^{n}_{j,k=1} \frac{\partial^{2}\sigma_{i}}{\partial v_{j}\partial v_{k}}(v_{0})y_{j}y_{k}+.......
 \end{equation}
 where $y$ is defined by $y=v-v_{0}$. Now we neglect the second order or above derivative terms and define the Jacobian matrix as
 \begin{equation}\label{22}
    J=\frac{\partial \sigma_{i}}{\partial v_{j}}=
   \begin{pmatrix}
   \frac{ \partial \sigma_{1}}{\partial v_{1}} &  \frac{ \partial \sigma_{1}}{\partial v_{2}} & \cdots &  \frac{ \partial \sigma_{1}}{\partial v_{n}} \\
   \vdots&  \cdots & \cdots & \vdots \\
     \frac{ \partial \sigma_{n}}{\partial v_{1}} &  \frac{ \partial \sigma_{n}}{\partial v_{2}} & \cdots &  \frac{ \partial \sigma_{n}}{\partial v_{n}}
  \end{pmatrix},
 \end{equation}
 This matrix is known as stability matrix. The eigenvalues are evaluated from Jacobian matrix $J$ for equilibrium point $v_{0}$. The equilibrium point $v_{0}$ is hyperbolic if the characteristic roots of the matrix $J$ have non zero real part, else the point $v_{0}$ is non-hyperbolic. If all the characteristic roots of the matrix $J$ are positive and trajectories are move away from the point then the fixed point $v_{0}$ is known as unstable point or repeller. If all the characteristic roots are negative and the point attracts all nearby trajectories then the equilibrium point $v_{0}$ is known as stable as well as attractor. If two eigenvalues both are opposite in signs with real part and trajectories attract in some directions and repels along other directions then the equilibrium point $v_{0}$ is known as saddle node. \\
 \subsection{Model 1: $G(T)$ = $\alpha T+\frac{\beta}{T}$}\label{sec3.1}
 We take the form of $G(T)$ as \cite{Myrzakulov11}
 \begin{equation}\label{23}
   G(T) = \alpha T+\frac{\beta}{T},
 \end{equation}
 where $\alpha$ and $\beta$ are the free parameters. By using equation (\ref{23}), the expressions of pressure and energy density for dark energy are as follows:
 \begin{equation}\label{24}
   \rho_{de}=\frac{\dot{\phi}^{2}}{2}+V(\phi)+6\alpha H^{2}-\frac{\beta}{2H^{2}}\; ,
 \end{equation}
 \begin{equation}\label{25}
   p_{de}=\frac{\dot{\phi}^{2}}{2}-V(\phi)-6\alpha H^{2}+\frac{\beta}{2H^{2}}-4\dot{H}\bigg(\alpha+\frac{\beta}{12H^{4}}\bigg)\; ,
 \end{equation}
 and Klein-Gordon equation (\ref{13}) can be expressed as,
 \begin{equation}\label{26}
   \ddot{\phi}+3H\dot{\phi}+V_{,\phi}=0 ,
 \end{equation}
 By using equations (\ref{24}-\ref{25}), the EoS parameter is as follows:
 \begin{equation}\label{27}
   \omega_{de}=\frac{p_{de}}{\rho_{de}}=\frac{\frac{\dot{\phi}^{2}}{2}-V(\phi)-6\alpha H^{2}+\frac{\beta}{2H^{2}}-4\dot{H}\bigg(\alpha+\frac{\beta}{12H^{4}}\bigg)}{\frac{\dot{\phi}^{2}}{2}+V(\phi)+6\alpha H^{2}-\frac{\beta}{2H^{2}}}.
 \end{equation}
 To discuss the stability behavior of the model, we assume new dimensionless variables from equation (\ref{24}) to find the set of differential equations as follows:
 \begin{equation}\label{28}
   x=\frac{k\dot{\phi}}{\sqrt{6}H}, \hspace{0.3cm} y=\frac{k\sqrt{V}}{\sqrt{3}H}, \hspace{0.3cm} z=2\alpha k^{2}, \hspace{0.3cm} r=-\frac{k^{2}\beta}{6H^{4}}
 \end{equation}
 \begin{equation}\label{29}
   \rho=\frac{k\sqrt{\rho_{r}}}{\sqrt{3}H}, \hspace{0.3cm} \lambda=-\frac{V'(\phi)}{kV(\phi)}, \hspace{0.3cm} \sigma=\frac{V''(\phi)V(\phi)}{V'(\phi)^{2}}.
 \end{equation}
 where $k^{2}=1$. The density parameters with regard to the dimensionless variables are obtained as,
 \begin{equation}\label{30}
   \Omega_{r}=\frac{k^{2}\rho_{r}}{3H^{2}}=\rho^{2},
 \end{equation}
 \begin{equation}\label{31}
   \Omega_{m}=\frac{k^{2}\rho_{m}}{3H^{2}}=1-x^{2}-y^{2}-z-r-\rho^{2},
 \end{equation}
 \begin{equation}\label{32}
   \Omega_{de}=\frac{k^{2}\rho_{de}}{3H^{2}}=x^{2}+y^{2}+z+r.
 \end{equation}
 From the field equations of $f(T,\phi)$ gravity (\ref{11}-\ref{12}) using the above variables of equation (\ref{28}), we will get
 \begin{equation}\label{33}
   \frac{\dot{H}}{H^{2}}=\frac{\rho^{2}-3(r-x^{2}+y^{2}+z-1)}{2z-2r-2},
 \end{equation}
 By using equation (\ref{33}), the expressions of deceleration parameter and the EoS parameter are obtained as,
 \begin{equation}\label{34}
   q=-1-\frac{\dot{H}}{H^{2}}=\frac{\rho^{2}-5r+3x^{2}-3y^{2}-z+1}{2r-2z+2},
 \end{equation}
 \begin{equation}\label{35}
   \omega_{de}=\frac{-2\rho^{2}(z-r)-6x^{2}+6y^{2}+12r}{3(x^{2}+y^{2}+r+z)(2z-2r-2)},
 \end{equation}
 By using the relation of the deceleration parameter and the EoS parameter $(\omega_{tot})$, the expression of $\omega_{tot}$ in terms of the variables is
 \begin{equation}\label{36}
  \omega_{tot} =\frac{2 q-1}{3}=\frac{\rho^{2}-6r+3x^{2}-3y^{2}}{3r-3z+3}.
 \end{equation}
 From the variables in equation (\ref{28}), the set of differential equations can be derived as,
 \begin{equation}\label{37}
   x'=\sqrt{\frac{3}{2}}y^{2}\lambda-3x-\frac{x[\rho^{2}-3(r-x^{2}+y^{2}+z-1)]}{2z-2r-2},
 \end{equation}
 \begin{equation}\label{38}
   y'=-\sqrt{\frac{3}{2}}xy\lambda-\frac{y[\rho^{2}-3(r-x^{2}+y^{2}+z-1)]}{2z-2r-2},
 \end{equation}
 \begin{equation}\label{39}
   z'=0,
 \end{equation}
 \begin{equation}\label{40}
   r'=\frac{-4r[\rho^{2}-3(r-x^{2}+y^{2}+z-1)]}{2z-2r-2},
 \end{equation}
 \begin{equation}\label{41}
   \rho'=\frac{-\rho(\rho^{2}-7r+3x^{2}-3y^{2}+z-1)}{2z-2r-2},
 \end{equation}
 \begin{equation}\label{42}
   \lambda'=-\sqrt{6}x(\sigma-1)\lambda^{2},
 \end{equation}
 We discuss the stability behavior of the above system of equations (\ref{37}-\ref{42}) at each equilibrium points with the scalar potential function $V(\phi)=V_{0}e^{-\lambda \phi}$. To evaluate the equilibrium points for the system of equations (\ref{37}-\ref{42}), we solve the equations $x'=y'=z'=r'=\rho'=\lambda'=0$. We found eight equilibrium points which are as follows\\
 (1) Equilibrium Point $(A)$: $x=0$, $y=0$, $z=0$, $r=0$, $\rho=0$,\\
 (2) Equilibrium Point $(B)$: $x=0$, $y=0$, $z=0$, $r=1$, $\rho=0$,\\
 (3) Equilibrium Point $(C)$: $x=0$, $y=0$, $z=\nu$, $r=\mu$, $\rho=0$ where $\mu=1-\nu$, $\nu\neq0$,\\
 (4) Equilibrium Point $(D)$: $x=\delta$, $y=0$, $z=0$, $r=\gamma$, $\rho=0$ where $\gamma=1-\delta^{2}$, $\delta\neq0$,\\
 (5) Equilibrium Point $(E)$: $x=\frac{\sqrt{\frac{3}{2}}}{\lambda}$, $y=\frac{1}{\lambda}\sqrt{\frac{3}{2}}$, $z=0$, $r=0$, $\rho=0$ where $\lambda\neq0$,\\
 (6) Equilibrium Point $(F)$: $x=0$, $y=\zeta$, $z=0$, $r=\epsilon$, $\rho=0$ where $\epsilon=1-\zeta^{2}$, $\zeta\neq0$,\\
 (7) Equilibrium Point $(G)$: $x=0$, $y=\tau$, $z=\eta$, $r=\chi$, $\rho=0$ where $\chi=1-\tau^{2}-\eta^{2}$,\\
 (8) Equilibrium Point $(H)$: $x=\frac{\lambda}{\sqrt{6}}$, $y=\sqrt{1-\frac{\lambda^{2}}{6}}$, $z=0$, $r=0$, $\rho=0$.\\

 The characteristic values are evaluated by solving the Jacobian matrix at each equilibrium points presented in Table 1. The density parameters, deceleration and EoS parameters for the equilibrium points are presented in Table 2.\\

 \begin{table}[hbt!]
\begin{center}
\caption{Characteristic values and stability conditions for the equilibrium points of the system of equations (\ref{37}-\ref{42})}
\begin{tabular}{|p{2cm}|p{2cm}|p{2cm}|p{2cm}|p{2cm}|p{4cm}|}
\hline
Points & $\hspace{0.3cm}\lambda_{1}$ & $\hspace{0.3cm}\lambda_{2}$ & $\hspace{0.3cm}\lambda_{3}$ & $\hspace{0.5cm}\lambda_{4}$ & Nature\\
\hline
$A$ & $\hspace{0.3cm}-\frac{9}{2}$ & $\hspace{0.3cm}\frac{3}{2}$ & $\hspace{0.3cm}0$ & $\hspace{0.3cm}-\frac{1}{2}$ & unstable saddle\\
\hline
$B$ & $\hspace{0.3cm}-3$ & $\hspace{0.3cm}0$ & $\hspace{0.1cm}-3$ & $\hspace{0.3cm}-2$ & stable\\
\hline
$C$ & $\hspace{0.3cm}-3$ & $\hspace{0.1cm}-3$ & $\hspace{0.3cm}0$ & $\hspace{0.3cm}-2$ & stable\\
\hline
$D$ & $\hspace{0.3cm}\frac{6-9\delta^{2}}{\delta^{2}-2}$ & $\frac{6+\sqrt{6}\delta\lambda}{2}$ & $\hspace{0.3cm}\frac{6(1-3\delta^{2})}{\delta^{2}-2}$ & $\hspace{0.2cm}\frac{4-5\delta^{2}}{\delta^{2}-2}$ & unstable at $\delta=1$\\
\hline
$E$ & $\hspace{0.3cm}-\frac{3}{2}$ & $\hspace{0.3cm}-\frac{9}{2\lambda^{2}}$ & $\hspace{0.3cm}0$ & $\hspace{0.3cm}-\frac{1}{2}$ & stable at $\lambda^{2}=3$\\
\hline
$F$ & $\hspace{0.3cm}-3$ & $\hspace{0.2cm}\frac{3\zeta^{2}}{\zeta^{2}-2}$ & $\hspace{0.2cm}-\frac{6(1-\zeta^{2}))}{\zeta^{2}+2}$ & $\hspace{0.3cm}-2$ & stable at $\zeta^{2}=1$\\
\hline
$G$ & $\hspace{0.3cm}-3$ & $\hspace{0.2cm}\frac{3\tau^{2}}{\tau^{2}+2\eta-2}$ & $\hspace{0.2cm}-\frac{6(\tau^{2}+\eta-1))}{\tau^{2}+2\eta-2}$ & $\hspace{0.3cm}-2$ & stable at $\tau^{2}=2$, $\eta   =-1$\\
\hline
$H$ & $\hspace{0.3cm}\lambda^{2}-3$ & $\hspace{0.3cm}\frac{\lambda^{2}-6}{2}$ & $\hspace{0.3cm}2\lambda^{2}$ & $\hspace{0.3cm}\frac{\lambda^{2}-4}{2}$ & stable for $\lambda^{2}<0$, unstable for $\lambda^{2}>0$\\
\hline
\end{tabular}
\end{center}
\end{table}

\begin{table}[hbt!]
\begin{center}
\caption{Density parameters, EoS and deceleration parameters at these points}
\begin{tabular}{ |p{1cm}|p{1cm}|p{1.7cm}|p{0.6cm}|p{1.6cm}|p{1.6cm}|p{0.9cm}|}
\hline
Points &\hspace{0.4cm} $\Omega_{m}$ &\hspace{0.5cm} $\Omega_{r}$ & $\Omega_{de}$ & $\hspace{0.5cm}\omega_{de}$ &\hspace{0.5cm} $\omega_{tot}$ & $\hspace{0.2cm} q$ \\
\hline
$A$ & $\hspace{0.5cm}1$ & $\hspace{0.6cm}0$ & $0$ & $\hspace{0.6cm}0$ & $\hspace{0.6cm}0$ & $\hspace{0.2cm}\frac{1}{2}$ \\
\hline
$B$ & $\hspace{0.5cm}0$ & $\hspace{0.6cm}0$ & $1$ & $\hspace{0.5cm}-1$ & $\hspace{0.5cm}-1$ & $\hspace{0.1cm}-1$ \\
\hline
$C$ & $\hspace{0.5cm}0$ & $\hspace{0.6cm}0$ & $1$ & $\hspace{0.5cm}-1$ & $\hspace{0.5cm}-1$ & $\hspace{0.1cm}-1$ \\
\hline
$D$ & $\hspace{0.5cm}0$ & $\hspace{0.6cm}0$ & $1$ & $\hspace{0.6cm}1$ & $\hspace{0.7cm}1$ & $\hspace{0.3cm}2$ \\
\hline
$E$ & $\hspace{0.5cm}0$ & $\hspace{0.6cm}0$ & $1$ & $\hspace{0.6cm}0$ & $\hspace{0.6cm}0$ & $\hspace{0.2cm}\frac{1}{2}$ \\
\hline
$F$ & $\hspace{0.5cm}0$ & $\hspace{0.6cm}0$ & $1$ & $\hspace{0.5cm}-1$ & $\hspace{0.5cm}-1$ & $\hspace{0.1cm}-1$ \\
\hline
$G$ & $\hspace{0.5cm}0$ & $\hspace{0.6cm}0$ & $1$ & $\hspace{0.5cm}-1$ & $\hspace{0.5cm}-1$ & $\hspace{0.1cm}-1$ \\
\hline
$H$ & $\hspace{0.5cm}0$ & $\hspace{0.6cm}0$ & $1$ & $\hspace{0.5cm}-1$ & $\hspace{0.5cm}-1$ & $\hspace{0.1cm}-1$ \\
\hline
\end{tabular}
\end{center}
\end{table}
\begin{figure}[hbt!]
\begin{center}
  \includegraphics[scale=0.8]{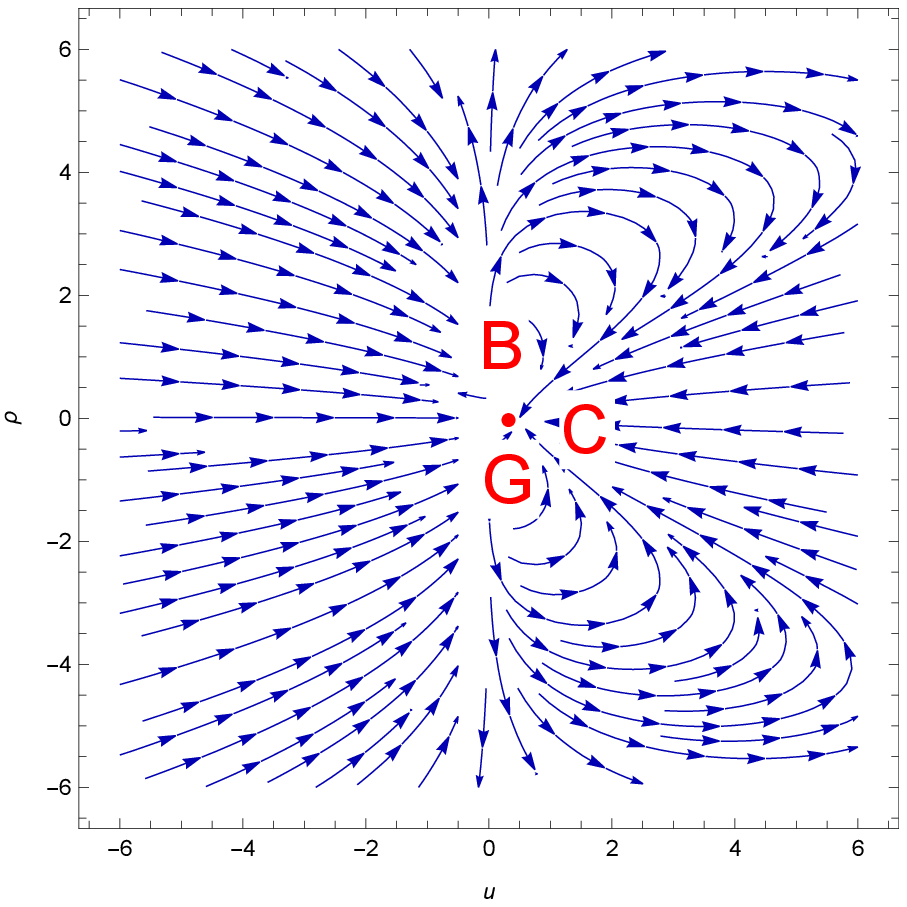}~~~~~~~~~
  \includegraphics[scale=0.8]{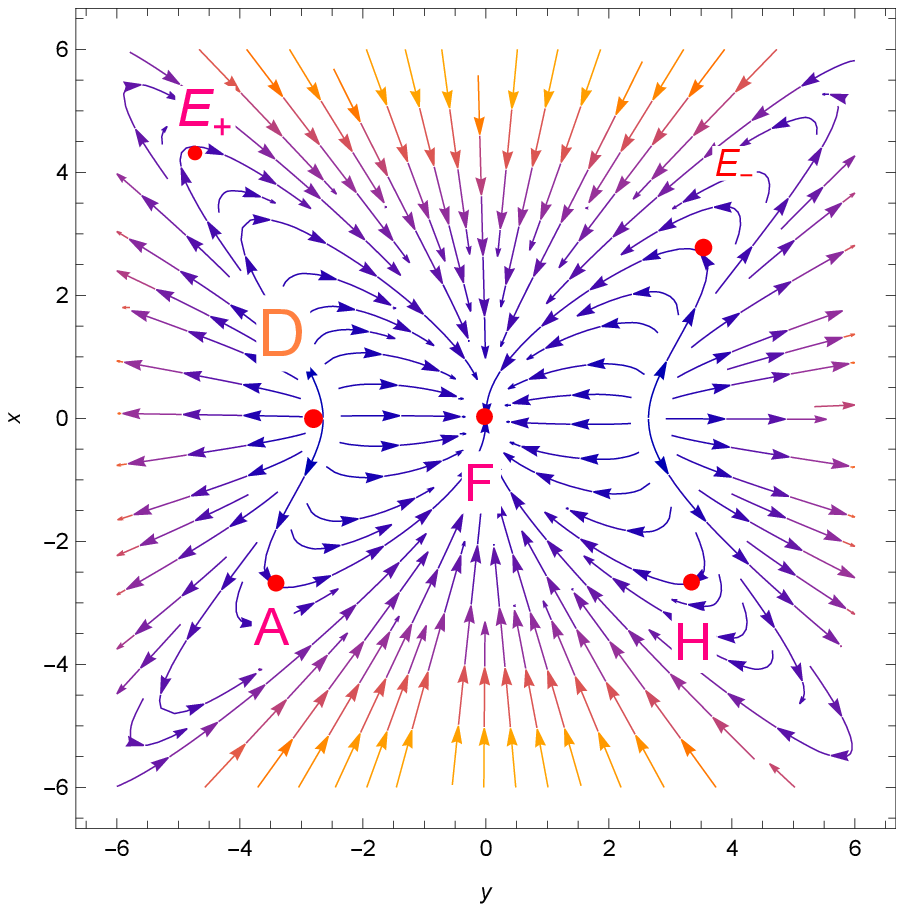}~~~~\\
{\bf{Figure. I:}} Phase plot for the system of equations (\ref{37}-\ref{42}), $(i)$ \textbf{in left plot}($z=0.5$, $\lambda=0.005$), $(ii)$ \textbf{in right plot}($\lambda=0.005$, $z=0.5$, $r=1.5$).

  \hspace{1cm}\vspace{5mm}

\vspace{3mm}
\vspace{3mm}
\end{center}
\end{figure}
From Table 1, for the point $A$, two characteristic values $\lambda_{1}$ and $\lambda_{4}$ are negative and one characteristic value $\lambda_{2}$ is positive. For both positive and negative characteristic values, the equilibrium point $A$ is saddle point and from Figure I (right plot), for this point, the trajectories diverge to the equilibrium point. From Table 2, at the point $A$, the density parameters $\Omega_{m}=1$ and $\Omega_{r}=\Omega_{de}=0$ represent the matter dominant stage of the Universe and $q=\frac{1}{2}$, EoS parameters $\omega_{de}=\omega_{tot}=0$ show the decelerated stage of the Universe. Again for the point $B$, the characteristic values are $(-3,0,-3,-2)$. Since all the characteristic values are negative so the point $B$ is stable point and from Figure I (left plot) for the point $B$ all the trajectories are directed towards the equilibrium point. Also, at this point the density parameters $\Omega_{m}=\Omega_{r}=0$ and $\Omega_{de}=1$ represent the dark energy dominant Universe and EoS parameters $\omega_{de}=\omega_{tot}=-1$, deceleration parameter $q=-1$ assure the accelerated phase of the Universe. For the point $C$, the characteristic values are $(-3,0,-3,-2)$. Since all the characteristic values are negative so the point $C$ is stable point and from Figure I (left plot) for the point $C$ all the trajectories are directed towards the equilibrium point. Also, at this point, the density parameters $\Omega_{m}=\Omega_{r}=0$ and $\Omega_{de}=1$ represent the dark energy dominant Universe and EoS parameters $\omega_{de}=\omega_{tot}=-1$, deceleration parameter $q=-1$ assure the accelerated phase of the Universe. Further, the characteristic values for the point $D$ are $\bigg(\frac{6-9\delta^{2}}{\delta^{2}-2},\frac{6+\sqrt{6}\delta\lambda}{2},\frac{6(1-3\delta^{2})}{\delta^{2}-2},\frac{4-5\delta^{2}}{\delta^{2}-2}\bigg)$. At $\delta=1$, the values are $(3,\frac{6+\sqrt{6}\lambda}{2},12,1)$. Since all the values are positive, so the point $D$ is unstable point and from Figure I (right plot) for this point $D$, the direction of all the trajectories move away from this point. The density parameters $\Omega_{m}=\Omega_{r}=0$, and $\Omega_{de}=1$ represent the dark energy dominant Universe at $\delta=1$, deceleration parameter $q=2$ and EoS parameters $\omega_{de}=\omega_{tot}=1$ assure the decelerated phase of the Universe. The characteristic values for the point $E$ are $(-\frac{3}{2},-\frac{9}{2\lambda^{2}},0,-\frac{1}{2})$. For $\lambda^{2}=3$, these values become $(-\frac{3}{2},-\frac{3}{2},0,-\frac{1}{2})$. Since all the characteristic values are negative so the point $E$ is stable node and Figure I (right plot) for the point $E$, all the trajectories are directed towards the equilibrium point. The density parameters are $\Omega_{m}=1-\frac{3}{\lambda^{2}}$, $\Omega_{de}=\frac{3}{\lambda^{2}}$, $\Omega_{r}=0$. At $\lambda^{2}=3$, the values becomes $\Omega_{m}=0$, $\Omega_{r}=0$ and $\Omega_{de}=1$ which represent the dark energy dominant Universe while deceleration parameter $q=\frac{1}{2}$, EoS parameters $\omega_{de}=\omega_{tot}=0$ represents the decelerated stage of the Universe. The characteristic values for the point $F$ are $\bigg(-3,\frac{3\zeta^{2}}{\zeta^{2}-2},-\frac{6(1-\zeta^{2}))}{\zeta^{2}+2},-2\bigg)$. At $\zeta^{2}=1$, the values become $(-3,-3,0,-2)$. All the characteristic values are negative so the point $F$ is stable node. $\Omega_{m}=\Omega_{r}=0$, and $\Omega_{de}=1$ represents the dark energy dominant Universe, the deceleration parameter $q=-1$, and EoS parameters $\omega_{de}=\omega_{tot}=-1$ assure the accelerated phase of the Universe. The characteristic values for the point $G$ are $\bigg(-3,\frac{3\tau^{2}}{\tau^{2}+2\eta-2},-\frac{6(\tau^{2}+\eta-1))}{\tau^{2}+2\eta-2},-2\bigg)$. At $\tau^{2}=2$, $\eta =-1$ the values are becomes $(-3,-3,0,-2)$. All the characteristic values are negative so the point $G$ is stable node and Figure I (right plot) for the point $G$, all the trajectories are directed towards the equilibrium point. $\Omega_{m}=\Omega_{r}=0$ and $\Omega_{de}=1$ represent the dark energy dominant Universe, the deceleration parameter $q=-1$, EoS parameters $\omega_{de}=\omega_{tot}=-1$ assure the accelerated phase of the Universe. The characteristic values for the point $H$ are $\bigg(\lambda^{2}-3,\frac{\lambda^{2}-6}{2},2\lambda^{2},\frac{\lambda^{2}-4}{2}\bigg)$ and for $\lambda^{2}<0$, all the characteristic values are negative and for $\lambda^{2}>0$, all the characteristic values are positive. Thus, the point $H$ is stable for $\lambda^{2}<0$ and unstable saddle for $\lambda^{2}>0$. The density parameters $\Omega_{m}=\Omega_{r}=0$, $\Omega_{de}=1$ represents the dark energy dominant Universe, the deceleration parameter $q=-1$, EoS parameters $\omega_{de}=\omega_{tot}=-1$ assure the accelerated phase of the Universe.\\

\subsection{Model 2: $G(T)$ = $\zeta T$ ln$(\psi T)$}\label{sec3.2}
We assume the form of $G(T)$ as \cite{Setare12},
\begin{equation}\label{43}
  G(T) = \zeta T\; ln (\psi T),
\end{equation}
 where $\zeta$ and $\psi$ are the free parameters. By using equation (\ref{43}), the expressions of pressure and energy density for dark energy are as follows:
 \begin{equation}\label{44}
   \rho_{de}=\frac{\dot{\phi}^{2}}{2}+V(\phi)+2\zeta T+\zeta T\; ln (\psi T)\; ,
 \end{equation}
 \begin{equation}\label{45}
   p_{de}=\frac{\dot{\phi}^{2}}{2}-V(\phi)-12\zeta H^{2}-6\zeta H^{2}ln (6\psi H^{2})-4\dot{H}\bigg(\zeta ln (6\psi H^{2})+3\zeta\bigg)\; ,
 \end{equation}
 By using equations (\ref{44}-\ref{45}), the EoS parameter is as follows:
 \begin{equation}\label{46}
   \omega_{de}=\frac{p_{de}}{\rho_{de}}=\frac{\frac{\dot{\phi}^{2}}{2}-V(\phi)-12\zeta H^{2}-6\zeta H^{2}ln (6\psi H^{2})-4\dot{H}\bigg(\zeta ln (6\psi H^{2})+3\zeta\bigg)}{\frac{\dot{\phi}^{2}}{2}+V(\phi)+2\zeta T+\zeta T\; ln (\psi T)}.
 \end{equation}
 To discuss the stability behavior of the model, we assume new dimensionless variables from equation (\ref{44}) to find the set of differential equations are as follows:
 \begin{equation}\label{47}
   x=\frac{k\dot{\phi}}{\sqrt{6}H}, \hspace{0.3cm} y=\frac{k\sqrt{V}}{\sqrt{3}H}, \hspace{0.3cm} z=4\zeta k^{2}, \hspace{0.3cm} r=2\zeta ln (6\psi H^{2})k^{2}
 \end{equation}
 \begin{equation}\label{48}
   \rho=\frac{k\sqrt{\rho_{r}}}{\sqrt{3}H}, \hspace{0.3cm} \lambda=-\frac{V'(\phi)}{kV(\phi)}, \hspace{0.3cm} \sigma=\frac{V''(\phi)V(\phi)}{V'(\phi)^{2}}.
 \end{equation}
 where $k^{2}=1$. The density parameters with regard to the dimensionless variables are obtained as,
 \begin{equation}\label{49}
   \Omega_{r}=\frac{k^{2}\rho_{r}}{3H^{2}}=\rho^{2},
 \end{equation}
 \begin{equation}\label{50}
   \Omega_{m}=\frac{k^{2}\rho_{m}}{3H^{2}}=1-x^{2}-y^{2}-z-r-\rho^{2},
 \end{equation}
 \begin{equation}\label{51}
   \Omega_{de}=\frac{k^{2}\rho_{de}}{3H^{2}}=x^{2}+y^{2}+z+r.
 \end{equation}
 From the field equations of $f(T,\phi)$ gravity (\ref{11}-\ref{12}) along with the above variables of equation (\ref{47}), we will obtain
 \begin{equation}\label{52}
   \frac{\dot{H}}{H^{2}}=\frac{-\rho^{2}+3(r-x^{2}+y^{2}+z-1)}{2-3z-2r},
 \end{equation}
 By using equation (\ref{52}), the expression of deceleration parameter and the EoS parameter are obtained as,
 \begin{equation}\label{53}
   q=-1-\frac{\dot{H}}{H^{2}}=\frac{\rho^{2}-r+3x^{2}-3y^{2}+1}{2-3z-2r},
 \end{equation}
 \begin{equation}\label{54}
   \omega_{de}=\frac{6y^{2}-6x^{2}-3z-\rho^{2}(3z-2r)}{3(x^{2}+y^{2}+r+z)(3z+2r-2)},
 \end{equation}
 By using the relation of the deceleration parameter and the EoS parameter $(\omega_{tot})$, the expression of $\omega_{tot}$ in terms of the variables is
 \begin{equation}\label{55}
  \omega_{tot} =\frac{2 q-1}{3}=\frac{6x^{2}-6y^{2}+3z+2\rho^{2}}{6-9z-6r}.
 \end{equation}
  From the variables in equation (\ref{47}), the set of differential equations can be derived as,
 \begin{equation}\label{56}
   x'=\sqrt{\frac{3}{2}}y^{2}\lambda-3x-\frac{x[\rho^{2}-3(r-x^{2}+y^{2}+z-1)]}{3z+2r-2},
 \end{equation}
 \begin{equation}\label{57}
   y'=-\sqrt{\frac{3}{2}}xy\lambda-\frac{y[\rho^{2}-3(r-x^{2}+y^{2}+z-1)]}{3z+2r-2},
 \end{equation}
 \begin{equation}\label{58}
   z'=0,
 \end{equation}
 \begin{equation}\label{59}
   r'=\frac{\rho^{2}z-3z(r-x^{2}+y^{2}+z-1)]}{3z+2r-2},
 \end{equation}
 \begin{equation}\label{60}
   \rho'=\frac{\rho(1-\rho^{2}-r-3x^{2}+3y^{2}-3z)}{3z+2r-2},
 \end{equation}
 \begin{equation}\label{61}
   \lambda'=-\sqrt{6}x(\sigma-1)\lambda^{2},
   \end{equation}
 We discuss the stability behavior of the above system of equations (\ref{56}-\ref{61}) at each equilibrium points with the scalar potential function $V(\phi)=V_{0}e^{-\lambda \phi}$. To evaluate the equilibrium points for the system of equations (\ref{56}-\ref{61}), we solve the equations $x'=y'=z'=r'=\rho'=\lambda'=0$. We found eight equilibrium points which are as follows\\
 (1) Equilibrium Point $(A_{1})$: $x=0$, $y=0$, $z=0$, $r=\eta_{1}$, $\rho=0$,\\
 (2) Equilibrium Point $(B_{1})$: $x=0$, $y=0$, $z=\eta_{2}$, $r=1-\eta_{2}$, $\rho=0$ where $\eta_{2}\neq0$,\\
 (3) Equilibrium Point $(C_{1})$: $x=\eta_{3}$, $y=0$, $z=0$, $r=1-\eta_{3}^{2}$, $\rho=0$ where $\eta_{3}\neq0$,\\
 (4) Equilibrium Point $(D_{1})$: $x=0$, $y=0$, $z=0$, $r=1-\eta_{4}^{2}$, $\rho=\eta_{4}$ where $\eta_{4}\neq0$,\\
 (5) Equilibrium Point $(E_{1})$: $x=\frac{\sqrt{\frac{3}{2}}}{\lambda}$, $y=\pm\frac{1}{\lambda}\sqrt{\frac{3}{2}}$, $z=0$, $r=\eta_{5}$, $\rho=0$ where $\lambda\neq0$, $\eta_{5}\neq1$,\\
 (6) Equilibrium Point $(F_{1})$: $x=0$, $y=\eta_{6}$, $z=0$, $r=1-\eta_{6}^{2}$, $\rho=0$ where $\eta_{6}\neq0$,\\
 (7) Equilibrium Point $(G_{1})$: $x=0$, $y=0$, $z=\eta_{7}$, $r=\eta_{8}$, $\rho=\eta_{9}$ where $\eta_{9}=\pm\sqrt{3\eta_{7}+3\eta_{8}-3}$,\\

 The characteristic values evaluated by solving the Jacobian matrix at each equilibrium points are presented in Table 3 whereas the density parameters, deceleration and EoS parameters for the equilibrium points are presented in Table 4.\\

  \begin{table}[hbt!]
\begin{center}
\caption{Characteristic values and stability conditions for the equilibrium points of the system of equations (\ref{56}-\ref{61})}
\begin{tabular}{|p{2cm}|p{2cm}|p{2cm}|p{2cm}|p{2cm}|p{4cm}|}
\hline
Points & $\hspace{0.3cm}\sigma_{1}$ & $\hspace{0.3cm}\sigma_{2}$ & $\hspace{0.3cm}\sigma_{3}$ & $\hspace{0.5cm}\sigma_{4}$ & Nature\\
\hline
$A_{1}$ & $\hspace{0.3cm}-\frac{3}{2}$ & $\hspace{0.3cm}\frac{3}{2}$ & $\hspace{0.3cm}0$ & $\hspace{0.3cm}-\frac{1}{2}$ & unstable saddle\\
\hline
$B_{1}$ & $\hspace{0.3cm}-3$ & $\hspace{0.3cm}0$ & $\hspace{0.1cm}-3$ & $\hspace{0.3cm}-2$ & stable\\
\hline
$C_{1}$ & $\hspace{0.5cm}1$ & $\hspace{0.1cm}3-\sqrt{\frac{3}{2}}\lambda\eta_{3}$ & $\hspace{0.3cm}3$ & $\hspace{0.5cm}0$ & unstable\\
\hline
$D_{1}$ & $\hspace{0.5cm}2$ & $-3$ & $\hspace{0.3cm}1$ & $\hspace{0.5cm}0$ & unstable saddle\\
\hline
$E_{1}$ & $\hspace{0.3cm}-\frac{5}{4}$ & $\hspace{0.3cm}\frac{3}{2}$ & $\hspace{0.3cm}0$ & $\hspace{0.3cm}-\frac{1}{2}$ & unstable saddle\\
\hline
$F_{1}$ & $\hspace{0.3cm}-3$ & $\hspace{0.1cm}-2$ & $\hspace{0.3cm}0$ & $\hspace{0.3cm}-3$ & stable\\
\hline
$G_{1}$ & $\hspace{0.3cm}-3$ & $\hspace{0.4cm}0$ & $\hspace{0.1cm}-3$ & $\hspace{0.3cm}-1$ & stable\\
\hline
\end{tabular}
\end{center}
\end{table}

\begin{table}[hbt!]
\begin{center}
\caption{Density parameters, EoS and deceleration parameters at these points}
\begin{tabular}{ |p{1cm}|p{1cm}|p{1.7cm}|p{0.6cm}|p{1.6cm}|p{1.6cm}|p{0.9cm}|}
\hline
Points &\hspace{0.4cm} $\Omega_{m}$ &\hspace{0.5cm} $\Omega_{r}$ & $\Omega_{de}$ & $\hspace{0.5cm}\omega_{de}$ &\hspace{0.5cm} $\omega_{tot}$ & $\hspace{0.2cm} q$ \\
\hline
$A_{1}$ & $\hspace{0.5cm}1$ & $\hspace{0.6cm}0$ & $0$ & $\hspace{0.5cm}0$ & $\hspace{0.5cm}0$ & $\hspace{0.2cm}\frac{1}{2}$ \\
\hline
$B_{1}$ & $\hspace{0.5cm}0$ & $\hspace{0.6cm}0$ & $1$ & $\hspace{0.3cm}-1$ & $\hspace{0.3cm}-1$ & $-1$ \\
\hline
$C_{1}$ & $\hspace{0.5cm}0$ & $\hspace{0.6cm}0$ & $1$ & $\hspace{0.5cm}1$ & $\hspace{0.5cm}1$ & $\hspace{0.2cm}2$ \\
\hline
$D_{1}$ & $\hspace{0.5cm}0$ & $\hspace{0.6cm}0$ & $1$ & $\hspace{0.5cm}\frac{1}{3}$ & $\hspace{0.5cm}\frac{1}{3}$ & $\hspace{0.2cm}1$ \\
\hline
$E_{1}$ & $\hspace{0.5cm}1$ & $\hspace{0.6cm}0$ & $0$ & $\hspace{0.5cm}0$ & $\hspace{0.5cm}0$ & $\hspace{0.2cm}\frac{1}{2}$ \\
\hline
$F_{1}$ & $\hspace{0.5cm}0$ & $\hspace{0.6cm}0$ & $1$ & $\hspace{0.3cm}-1$ & $\hspace{0.3cm}-1$ & $-1$ \\
\hline
$G_{1}$ & $\hspace{0.5cm}0$ & $\hspace{0.6cm}0$ & $1$ & $\hspace{0.3cm}-1$ & $\hspace{0.3cm}-1$ & $-1$ \\
\hline
\end{tabular}
\end{center}
\end{table}
\begin{figure}[hbt!]
\begin{center}
\begin{subfigure}[b]{0.3\textwidth}
  \includegraphics[scale=0.4]{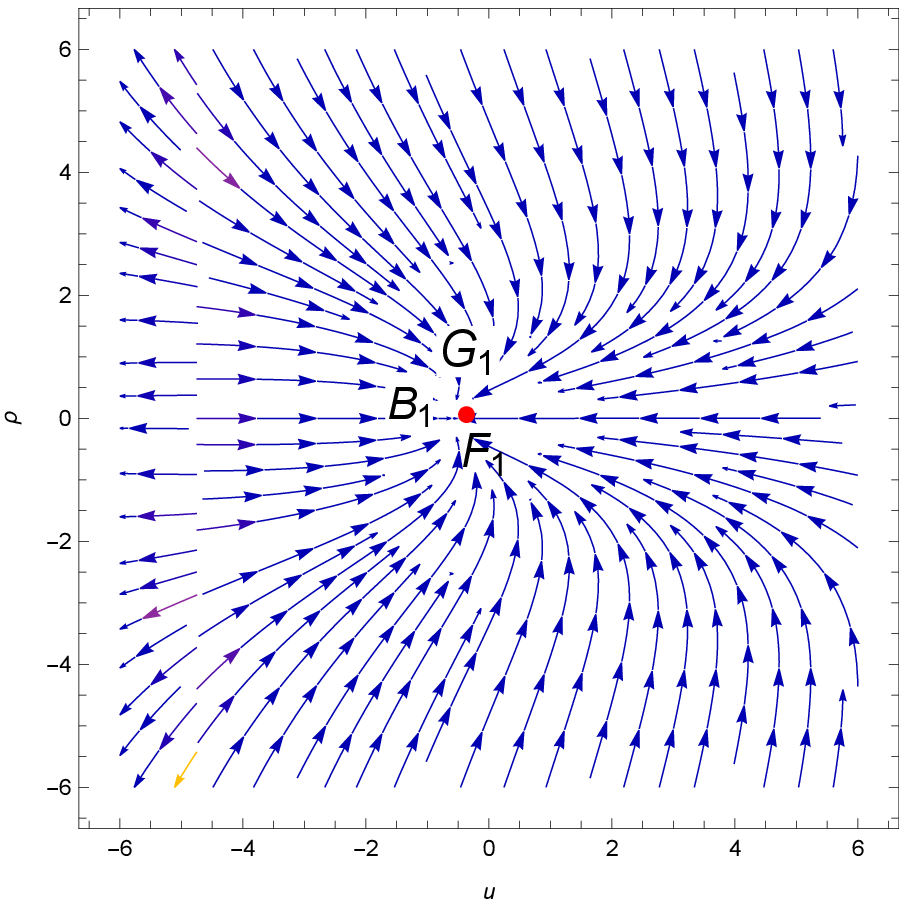}
  \end{subfigure}
   \hfill
     \begin{subfigure}[b]{0.3\textwidth}
  \includegraphics[scale=0.4]{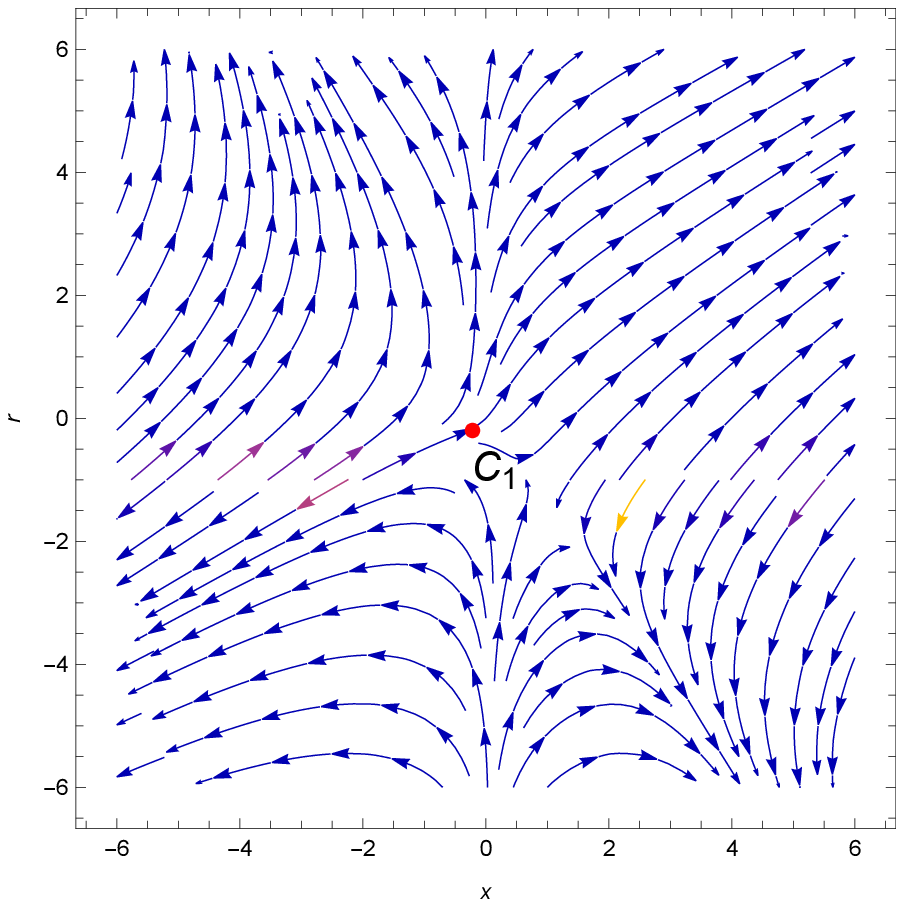}
  \end{subfigure}
  \hfill
     \begin{subfigure}[b]{0.3\textwidth}
  \includegraphics[scale=0.4]{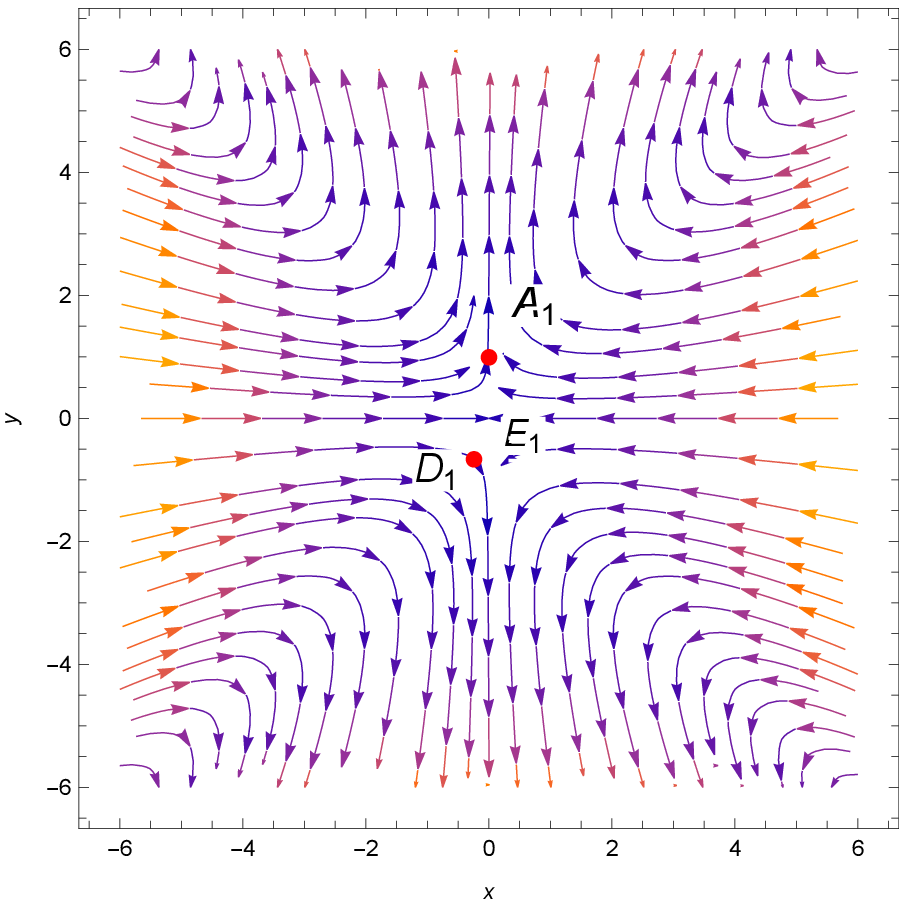}
  \end{subfigure}\\
  {\bf{Figure. II:}} Phase plot for the system of equations (\ref{56}-\ref{61}), $(i)$ \textbf{in left plot}($z=0.5$, $\lambda=0.005$), $(ii)$ \textbf{in middle plot}($\lambda=0.005$, $z=0.5$, $r=1.5$), $(iii)$ \textbf{in right plot}($y=0$, $\rho=0$).

  \hspace{1cm}\vspace{5mm}

\vspace{3mm}
\vspace{3mm}
\end{center}
\end{figure}
From Table 3, for the point $A_{1}$, two characteristic values $\sigma_{1}$ and $\sigma_{3}$ are negative and one characteristic value $\sigma_{2}$ is positive. For both positive and negative characteristic values, the equilibrium point $A_{1}$ is saddle point and from Figure II (middle plot), for this point, the trajectories diverge to the equilibrium point. From Table 4, at the point $A_{1}$ the density parameters $\Omega_{m}=1$, $\Omega_{r}=\Omega_{de}=0$ represent the matter dominant stage of the Universe and $q=\frac{1}{2}$, EoS parameters $\omega_{de}=\omega_{tot}=0$ shows the decelerated stage of the Universe. Again for the point $B_{1}$, the characteristic values are $(-3,0,-3,-2)$. Since all the characteristic values are negative so the point $B_{1}$ is stable point and from Figure II (left plot), for the point $B_{1}$, all the trajectories are directed towards the equilibrium point. Also, at this point, the density parameters $\Omega_{m}=\Omega_{r}=0$ and $\Omega_{de}=1$ represent the dark energy dominant Universe and EoS parameter $\omega_{de}=\omega_{tot}=-1$ and deceleration parameter $q=-1$ assure the accelerated phase of the Universe. For the point $C_{1}$, the characteristic values are $\bigg(1,3-\sqrt{\frac{3}{2}}\lambda\eta_{3},3,0\bigg)$. Since all the characteristic values are positive so the point $C_{1}$ is unstable point and from Figure II (right plot) for the point $C_{1}$ all the trajectories are directed towards the equilibrium point. Also, at this point, the density parameters $\Omega_{m}=\Omega_{r}=0$ and $\Omega_{de}=1$ represent the dark energy dominant Universe and EoS parameters $\omega_{de}=\omega_{tot}=1$, deceleration parameter $q=2$ assure the decelerated phase of the Universe. Further, the characteristic values for the point $D_{1}$ are $(-3,1,2,0)$. Since one characteristics value $\sigma_{1}$ is negative and other two values $\sigma_{2}$ and $\sigma_{3}$ are positive so the point $D_{1}$ is unstable saddle point and from Figure II (middle plot), for this point $D_{1}$, the direction of all the trajectories diverges to the equilibrium point. The density parameters $\Omega_{m}=\Omega_{r}=0$ and $\Omega_{de}=1$ represent the dark energy dominant Universe whereas deceleration parameter $q=1$ and EoS parameters $\omega_{de}=\omega_{tot}=\frac{1}{3}$ assure the decelerated phase of the Universe. The characteristic values for the point $E_{1}$ are $(-\frac{5}{4},\frac{3}{2},0,-\frac{1}{2})$. Since one characteristics value $\sigma_{2}$ is positive and other two values $\sigma_{1}$ and $\sigma_{3}$ are negative so the point $E_{1}$ is unstable saddle point and from Figure II (middle plot) for this point $E_{1}$, the direction of all the trajectories diverge to the equilibrium point. The density parameters are $\Omega_{m}=1$, $\Omega_{r}=0$ and $\Omega_{de}=0$ represent matter dominant stage of the Universe while deceleration parameter $q=\frac{1}{2}$, EoS parameters $\omega_{de}=\omega_{tot}=0$ represents the decelerated stage of the Universe. The characteristic values for the point $F_{1}$ are $(-3,-2,0,-3)$. All the characteristic values are negative so the point $F_{1}$ is stable node. $\Omega_{m}=\Omega_{r}=0$ and $\Omega_{de}=1$ represent the dark energy dominant Universe, the deceleration parameter $q=-1$ whereas EoS parameters $\omega_{de}=\omega_{tot}=-1$ assure the accelerated phase of the Universe. The characteristic values for the point $G_{1}$ are $(-3,0,-3,-1)$. All the characteristic values are negative so the point $G_{1}$ is stable node and Figure II (left plot) for the point $G_{1}$, all the trajectories are directed towards the equilibrium point. $\Omega_{m}=\Omega_{r}=0$ and $\Omega_{de}=1$ represent the dark energy dominant Universe, the deceleration parameter $q=-1$, EoS parameters $\omega_{de}=\omega_{tot}=-1$ assure the accelerated phase of the Universe.\\

\section{Hybrid Expansion Law in $f(T,\phi)$ gravity model}\label{sec4}
Several authors already use the power and exponential law form of scale factor $a(t)$ to establish the cosmological models which give detail explanation of the evolution history of the Universe. But these do not consider the transition phase from the early Universe and the accelerated phase of the present Universe this shows that the deceleration parameter $q=-1-\frac{\dot{H}}{H^{2}}$ is constant all over the cosmic evolution. Hybrid expansion law is one of the substitute way to solve this problem. We assume the relation between scale factor and Brans-Dicke scalar field is as follows \cite{Yadav21},
\begin{equation}\label{62}
  \phi=\phi_{0}a^{\gamma},
\end{equation}
where $\phi_{0}$ and $\gamma$ are constants. From current observations our present Universe belongs to the accelerated stage and at past it belongs to the decelerated stage. Thus, to find the proper model of the transition Universe,  we consider the scale factor $a(t)$ by the following hybrid expansion law \cite{Bhardwaj19,Tripathy20,Bennai22},
\begin{equation}\label{63}
  a=a_{1}t^{\delta}e^{\sigma t},
\end{equation}
where $a_{1}$, $\delta$ and $\sigma$ are real positive constants. By substituting the value of $a$ in equation (\ref{62}) we get,
\begin{equation}\label{64}
  \phi=\phi_{0}(a_{1}t^{\delta}e^{\sigma t})^{\gamma},
\end{equation}
\begin{figure}[hbt!]
\begin{center}
  \includegraphics[scale=0.5]{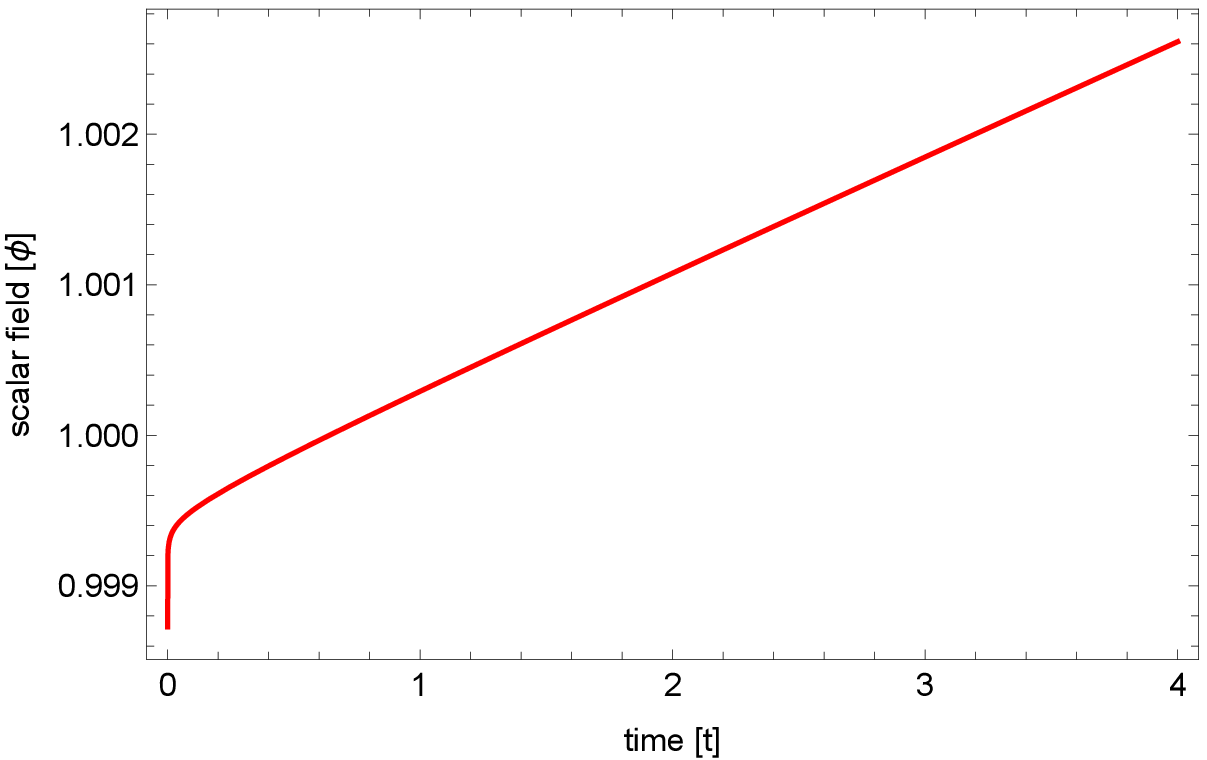}~~~~~~~~~

{\bf{Figure. III:}} Behavior of scalar field $\phi$ vs time $t$ for $a_{1}=0.4$, $\delta=0.7$, $\sigma=1.5$ and $\gamma=0.05$.
  \hspace{1cm}\vspace{5mm}

\vspace{3mm}
\vspace{3mm}

\end{center}
\end{figure}\\
Figure III represents the nature of the scalar field $(\phi)$ vs time $t$ by the choice of the parameters $a_{1}=0.4$, $\delta=0.7$, $\sigma=1.5$ and $\gamma=0.05$. The figure shows that $\phi$ increases over time. From equation (\ref{63}), the Hubble parameter can be evaluated as,
\begin{equation}\label{65}
   H= \frac{\dot{a}}{a}=\sigma+\frac{\delta}{t}.
\end{equation}
The deceleration parameter is derived by using the equation (\ref{65}) as,
\begin{equation}\label{66}
  q=-1-\frac{\dot{H}}{H^{2}}=-1+\frac{\delta}{(\delta+\sigma t)^{2}}.
\end{equation}
The important thing in cosmology is that to transform all the parameters in redshift. The equation of scale factor $(a(t))$ and the redshift parameter $(z)$ is $a(t)=\frac{a_{0}}{(1+z)}$ where $a_{0}=1$ be the current value of the scale factor. The relation between the time and redshift can be obtained as,
\begin{equation}\label{67}
  t(z)=\frac{\delta}{\sigma}W\bigg[\frac{\sigma}{\delta}\bigg(\frac{1}{a_{1}(1+z)}\bigg)^{\frac{1}{\delta}}\bigg],
\end{equation}
where $W$ represents the Lambert function which is also called "product algorithm".
\begin{figure}[hbt!]
\begin{center}
  \includegraphics[scale=0.5]{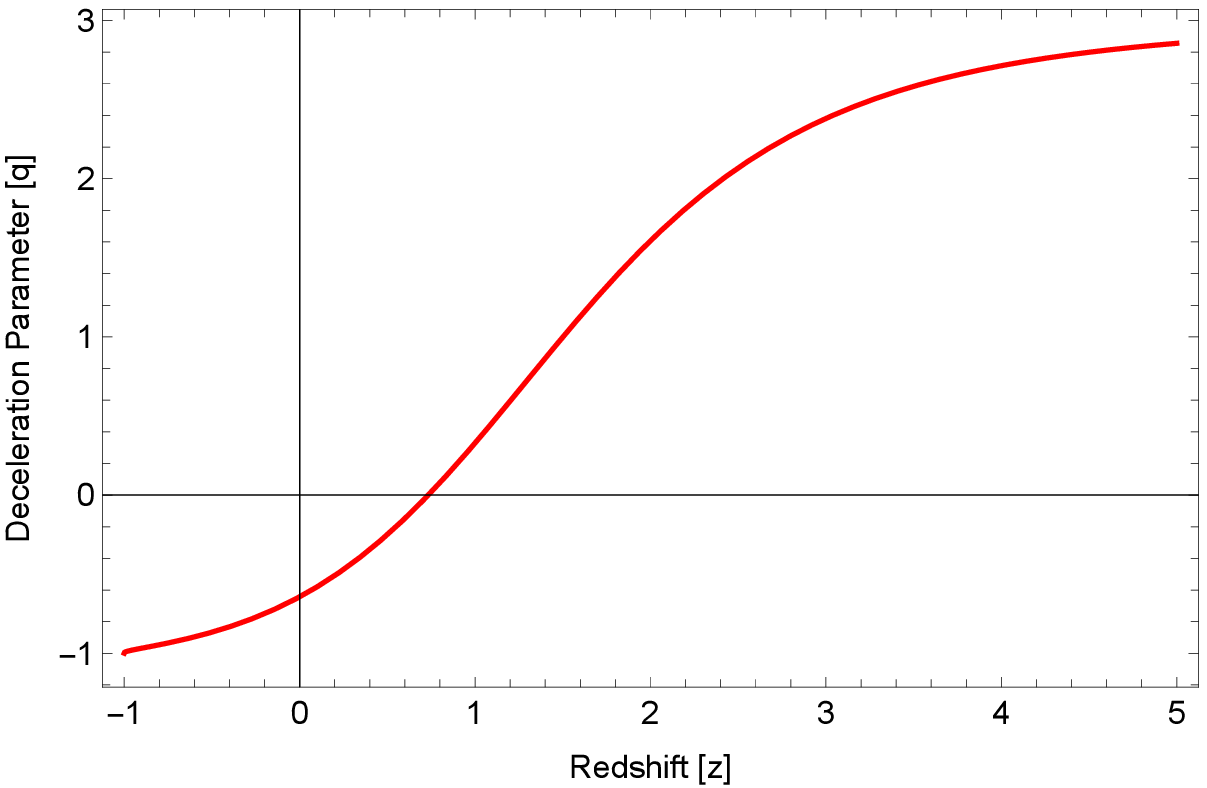}~~~~~~~~~

{\bf{Figure. IV:}} Plot of deceleration parameter $q$ vs redshift $z$ for $a_{1}=0.4$, $\delta=0.7$ and $\sigma=1.5$.
  \hspace{1cm}\vspace{5mm}

\vspace{3mm}
\vspace{3mm}

\end{center}
\end{figure}\\
The plot of deceleration parameter vs redshift is shown in Figure IV by the suitable choice of parameters. In Cosmology, the deceleration parameter explains the evolution history of the Universe from the decelerated phase of the early Universe to the accelerated phase of the current Universe. The Universe belongs to the decelerated stage at $q>0$, the expansion of the Universe is accelerated for $q<0$ and for $q=0$ the Universe is in marginal expansion. From current observations expansion of our present Universe is accelerated and the present value of the deceleration parameter belongs to the range $-1$ to $0$, $i.e.$ $-1\leq q\leq 0$. We write the deceleration parameter $(q)$ with regard to the redshift $z$ by using the equations (\ref{66}) and (\ref{67}). From Figure IV, we observe that for the parameters $a_{1}=0.4$, $\delta=0.7$ and $\sigma=1.5$, the deceleration parameter takes the positive value at initial stage of the Universe and it takes the negative value at present and finally it goes to $-1$ at $z=-1$. \\

For Model I, we assume the $G(T)$ function as $G(T)$ = $\alpha T+\frac{\beta}{T}$. By using this function in Friedmann equations (\ref{11}-\ref{12}), we evaluated the expressions of pressure and energy density as,
\begin{equation}\label{68}
  \rho_{de}=\frac{1}{2}\bigg[\phi_{0}^{2}\gamma^{2}a_{1}^{2\gamma}(t^{\delta}e^{\sigma t})^{2\gamma}\bigg(\frac{\delta}{t}+\sigma\bigg)^{2}\bigg]+V_{0}e^{-\lambda\phi_{0}[a_{1}t^{\delta}e^{\sigma t}]^{\gamma}}+6\alpha\bigg(\frac{\delta}{t}+\sigma\bigg)^{2}-\frac{\beta}{2\bigg(\frac{\delta}{t}+\sigma\bigg)^{2}},
\end{equation}
\begin{equation}\label{69}
  p_{de}=\frac{1}{2}\bigg[\phi_{0}^{2}\gamma^{2}a_{1}^{2\gamma}(t^{\delta}e^{\sigma t})^{2\gamma}\bigg(\frac{\delta}{t}+\sigma\bigg)^{2}\bigg]-V_{0}e^{-\lambda\phi_{0}[a_{1}t^{\delta}e^{\sigma t}]^{\gamma}}-6\alpha\bigg(\frac{\delta}{t}+\sigma\bigg)^{2}+\frac{\beta}{2\bigg(\frac{\delta}{t}+\sigma\bigg)^{2}}+\frac{4\delta}{t^{2}}\bigg(\alpha+\frac{\beta}{12\bigg(\frac{\delta}{t}+\sigma\bigg)^{4}}\bigg).
\end{equation}
\begin{figure}[hbt!]
\begin{center}
  \includegraphics[scale=0.5]{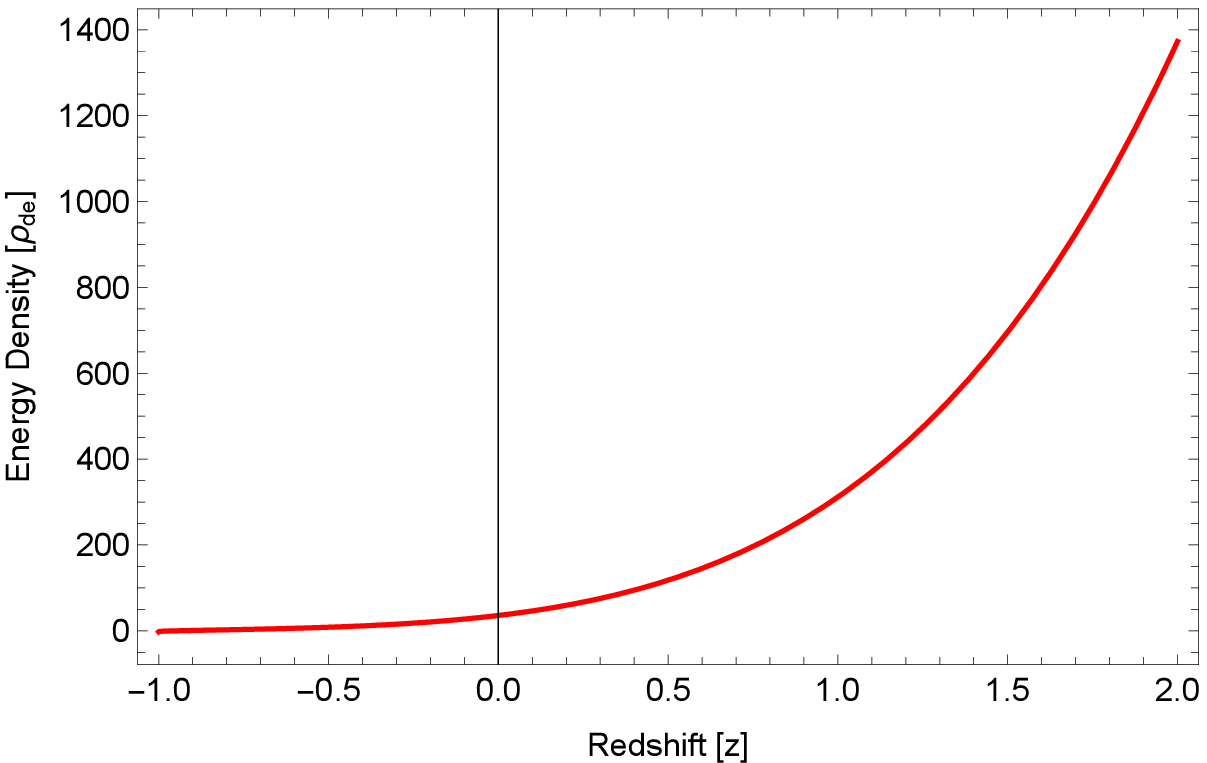}~~~~~~~~~
  \includegraphics[scale=0.5]{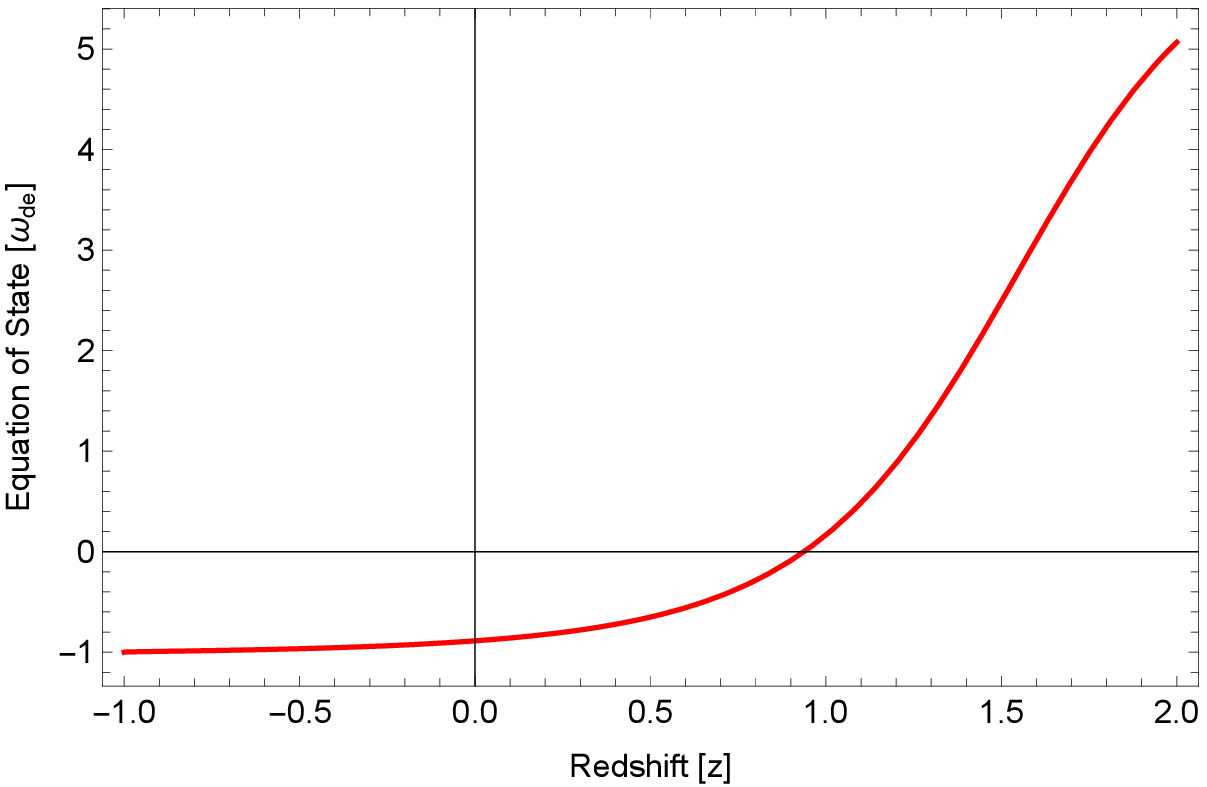}~~~~\\

{\bf{Figure. V:}} Behavior of energy density $(\rho)$ and pressure $(p)$ vs redshift $(z)$ for Model I with $a_{1}=0.4$, $\delta=0.7$, $\sigma=1.5$, $\gamma=0.05$, $\alpha=1.2$, $\beta=1.5$ and $\lambda=0.005$.
  \hspace{1cm}\vspace{5mm}

\vspace{3mm}
\vspace{3mm}

\end{center}
\end{figure}\\
To derive the above expressions, we use $\dot{\phi}=\phi_{0}\gamma a_{1}^{\gamma}(t^{\delta}e^{\sigma t})^{\gamma}\bigg(\frac{\delta}{t}+\sigma\bigg)$ and the scalar potential $V=V_{0}e^{-\lambda\phi_{0}[a_{1}t^{\delta}e^{\sigma t}]^{\gamma}}$. The nature of energy density vs redshift in $f(T,\phi)$ gravity Model I is given in Figure V. From Fig VI, we can say that energy density takes the positive values for all $z$ and the behavior of energy density function is increasing with redshift. At initial stage, the value of energy density is large positive number and finally it goes to zero for $z=-1$. The behavior of pressure vs redshift is given in Figure V. From the figure, we can say that for our model, the pressure $(p)$ starts with large positive number at initial stage and at the end, it goes to zero this means $p\rightarrow 0$ at $z\rightarrow-1$ in future. From current observations, our present Universe is in accelerated phase with positive energy density and negative pressure. We get the value of energy density is positive for all $z$ and pressure is negative in $f(T,\phi)$ gravity model which describes the acceleration of the present Universe.\\

For Model II, we assume the $G(T)$ function as $G(T)$ = $\zeta T$ ln$(\psi T)$. By using this function in Friedmann equations (\ref{11}-\ref{12}), we evaluated the expressions of pressure and energy density as,
\begin{equation}\label{70}
\rho_{de}=\frac{\dot{\phi}^{2}}{2}+V(\phi)+12\zeta\bigg(\frac{\delta}{t}+\sigma\bigg)^{2} +6\zeta\bigg(\frac{\delta}{t}+\sigma\bigg)^{2}ln\bigg[6\psi\bigg(\sigma+\frac{\delta}{t}\bigg)^{2}\bigg],
\end{equation}
\begin{equation}\label{71}
p_{de}=
\frac{\dot{\phi}^{2}}{2}-V(\phi)
-12\zeta\bigg(\frac{\delta}{t}+\sigma\bigg)^{2}-6\zeta\bigg(\frac{\delta}{t}+\sigma\bigg)^{2}ln\bigg[6\psi\bigg(\sigma+\frac{\delta}{t}\bigg)^{2}\bigg]+4\frac{\delta}{t^{2}}\bigg[\zeta ln\bigg[6\psi\bigg(\sigma+\frac{\delta}{t}\bigg)^{2}\bigg]+3\zeta\bigg],
\end{equation}

To derive the above expressions, we use $\dot{\phi}=\phi_{0}\gamma a_{1}^{\gamma}(t^{\delta}e^{\sigma t})^{\gamma}\bigg(\frac{\delta}{t}+\sigma\bigg)$ and the scalar potential $V=V_{0}e^{-\lambda\phi_{0}[a_{1}t^{\delta}e^{\sigma t}]^{\gamma}}$. The nature of energy density vs redshift in $f(T,\phi)$ gravity for Model II is given in Figure VI. From Fig. VI, we can say that energy density takes the positive values for all $z$ and the behavior of energy density function is increasing with redshift. At initial stage, the value of energy density is large positive number and finally it goes to zero for $z=-1$. The behavior of pressure vs redshift is given in Figure VI. From the Figure, we can say that for our model the pressure $(p)$ starts with large positive number at initial stage and at the end it goes to zero this means $p\rightarrow 0$ at $z\rightarrow-1$ in future. From current observations, our present Universe is in accelerated phase with positive energy density and negative pressure. We get the value of energy density is positive for all $z$ and pressure is negative in $f(T,\phi)$ gravity model which describes the acceleration of the present Universe.\\
\begin{figure}[hbt!]
\begin{center}
  \includegraphics[scale=0.5]{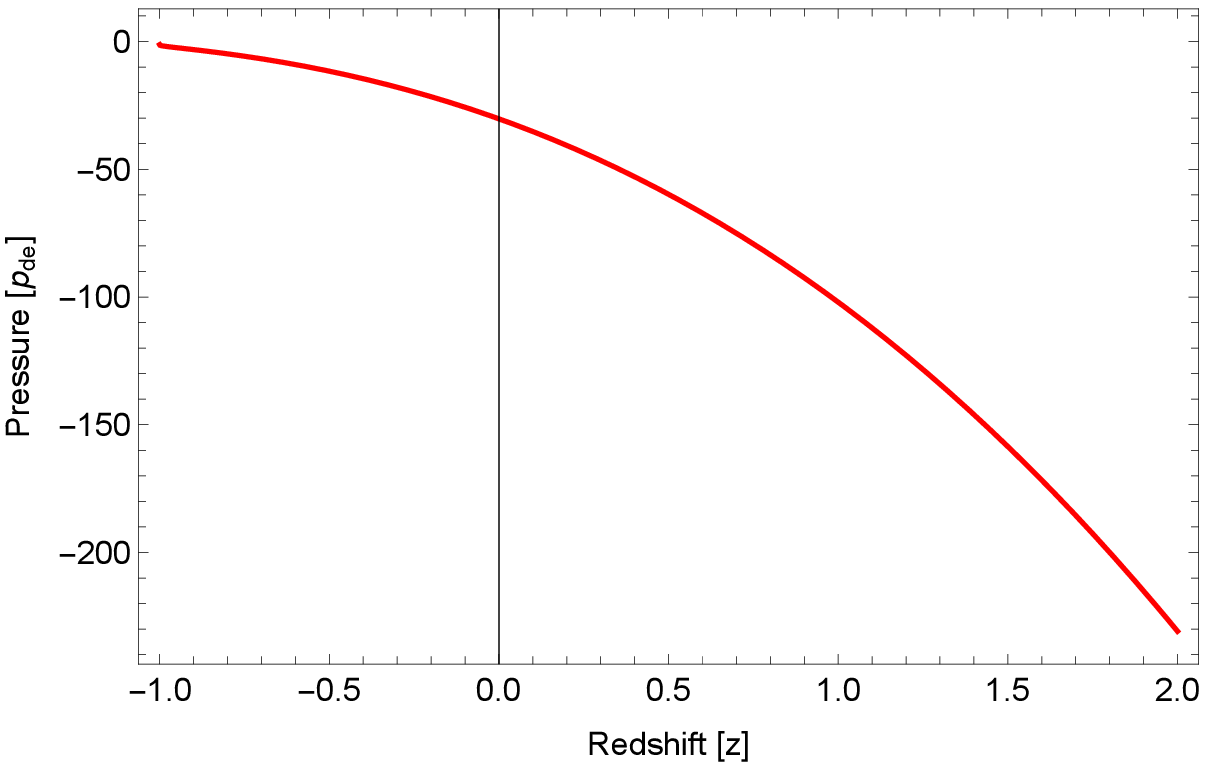}~~~~~~~~~
  \includegraphics[scale=0.5]{fig10.eps}~~~~\\

{\bf{Figure. VI:}} Behavior of energy density $(\rho)$ and pressure $(p)$ vs redshift $(z)$ for Model II with $a_{1}=0.4$, $\delta=0.7$, $\sigma=1.5$, $\gamma=0.05$, $\zeta=0.7$, $\psi=0.4$ and $\lambda=0.005$.
  \hspace{1cm}\vspace{5mm}

\vspace{3mm}
\vspace{3mm}

\end{center}
\end{figure}\\
For Model I, from equations (\ref{68}-\ref{69}), we evaluate the EoS parameter as,
\begin{equation}\label{72}
  \omega_{de}=\frac{p_{de}}{\rho_{de}}=\frac{\frac{\dot{\phi}^{2}}{2}-V(\phi)-6\alpha\bigg(\frac{\delta}{t}+\sigma\bigg)^{2}+\frac{\beta}{2\bigg(\frac{\delta}{t}+\sigma\bigg)^{2}}+\frac{4\delta}{t^{2}}\bigg(\alpha+\frac{\beta}{12\bigg(\frac{\delta}{t}+\sigma\bigg)^{4}}\bigg)}{\frac{\dot{\phi}^{2}}{2}+V(\phi)+6\alpha\bigg(\frac{\delta}{t}+\sigma\bigg)^{2}-\frac{\beta}{2\bigg(\frac{\delta}{t}+\sigma\bigg)^{2}}},
\end{equation}
For Model II, from equations (\ref{70}-\ref{71}), we evaluate the EoS parameter as,
\begin{equation}\label{73}
 \omega_{de}=\frac{\frac{\dot{\phi}^{2}}{2}-V(\phi)
-12\zeta\bigg(\frac{\delta}{t}+\sigma\bigg)^{2}-6\zeta\bigg(\frac{\delta}{t}+\sigma\bigg)^{2}ln\bigg[6\psi\bigg(\sigma+\frac{\delta}{t}\bigg)^{2}\bigg]+4\frac{\delta}{t^{2}}\bigg[\zeta ln\bigg[6\psi\bigg(\sigma+\frac{\delta}{t}\bigg)^{2}\bigg]+3\zeta\bigg]}{\frac{\dot{\phi}^{2}}{2}+V(\phi)+12\zeta\bigg(\frac{\delta}{t}+\sigma\bigg)^{2} +6\zeta\bigg(\frac{\delta}{t}+\sigma\bigg)^{2}ln\bigg[6\psi\bigg(\sigma+\frac{\delta}{t}\bigg)^{2}\bigg]},
\end{equation}
\begin{figure}[hbt!]
\begin{center}
  \includegraphics[scale=0.5]{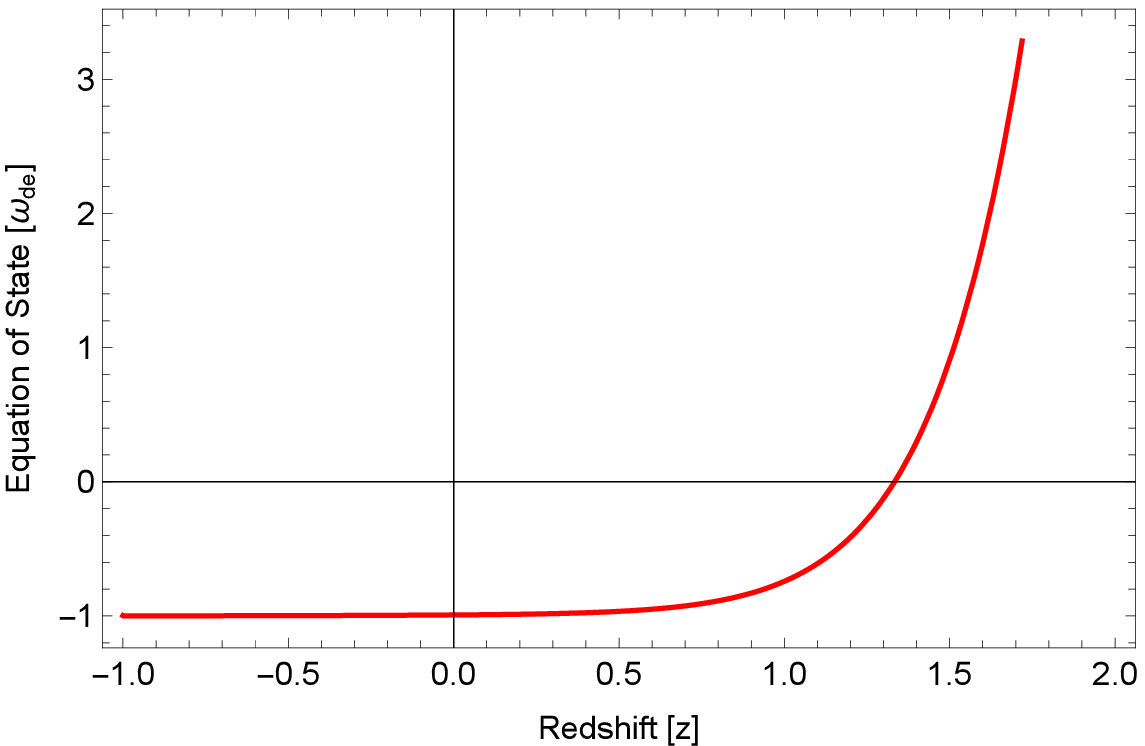}~~~~~~~~~
  \includegraphics[scale=0.5]{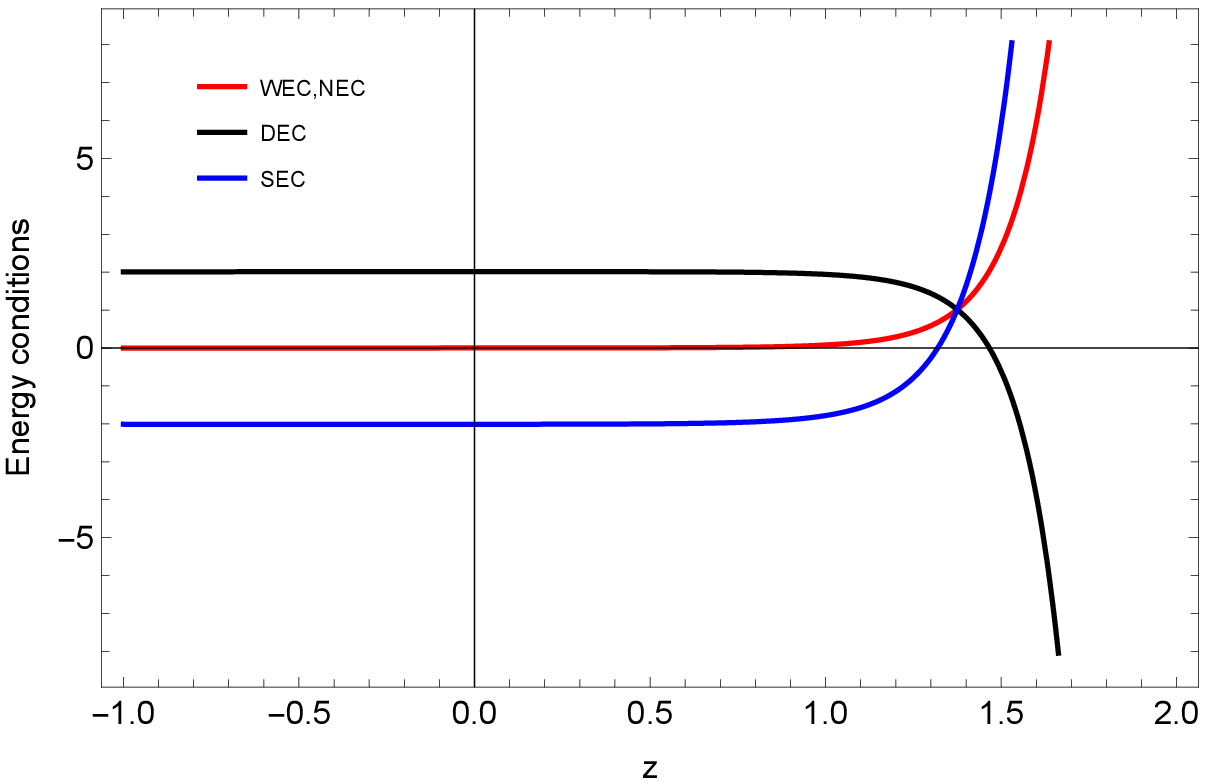}~~~~\\

{\bf{Figure. VII:}} Behavior of EoS parameter $(\omega_{de})$ vs redshift $(z)$ for Model I (left plot) with $a_{1}=0.4$, $\delta=0.7$, $\sigma=1.5$, $\gamma=0.05$, $\alpha=1.2$, $\beta=1.5$ and $\lambda=0.005$, for Model II (right plot) with $a_{1}=0.4$, $\delta=0.7$, $\sigma=1.5$, $\gamma=0.05$, $\zeta=0.7$, $\psi=0.4$ and $\lambda=0.005$.
  \hspace{1cm}\vspace{5mm}

\vspace{3mm}
\vspace{3mm}

\end{center}
\end{figure}\\
The nature of EoS parameter vs redshift is given in Figure VII. From current observations, the range of EoS parameter is $-1\leq \omega \leq 0$. If $\omega=1$ then it describes stiff fluid, for $-1<\omega\leq -\frac{1}{3}$ the Universe is in quintessence phase, if $\omega=-1$ then it represents $\Lambda$CDM model and for $\omega<-1$ it describes phantom model. From Fig. VII, we analyze that at $z=0$, $\omega$ belongs to the quintessence phase and $\omega$ approaches to $-1$ at $z=-1$ which represents the $\Lambda$CDM model. Thus from Fig VII, we observe that both the models describe the accelerating stage of the Universe. Also, for Model I we find the numerical value of EoS parameter is $\omega_{0}=-0.992$ and for Model II we find the numerical value of EoS parameter is $\omega_{0}=-0.883$ which satisfies the current Planck observational data.\\
\section{Energy Conditions}\label{sec5}
In Cosmology, the energy conditions contribute to a significant part to determine the prosperity of the Universe. These energy conditions are used to verified the acceleration epoch of the Universe. These energy conditions can be adopted by Raychaudhury equations in 1995. The types of the energy conditions in $f(T,\phi)$ gravity are \cite{Mishra22},\\
 $(1)$ Weak Energy Conditions (WEC): $\rho_{de}+p_{de}\geq 0$, $\rho_{de}\geq 0$, \\
 $(2)$ Null Energy Conditions (NEC): $\rho_{de}+p_{de}\geq 0$,\\
 $(3)$ Dominant Energy Conditions (DEC): $\rho_{de}-p_{de}\geq 0$,\\
 $(4)$ Strong Energy Conditions (SEC): $\rho_{de}+3p_{de}\geq 0$.\\
 \begin{figure}[hbt!]
\begin{center}
  \includegraphics[scale=0.5]{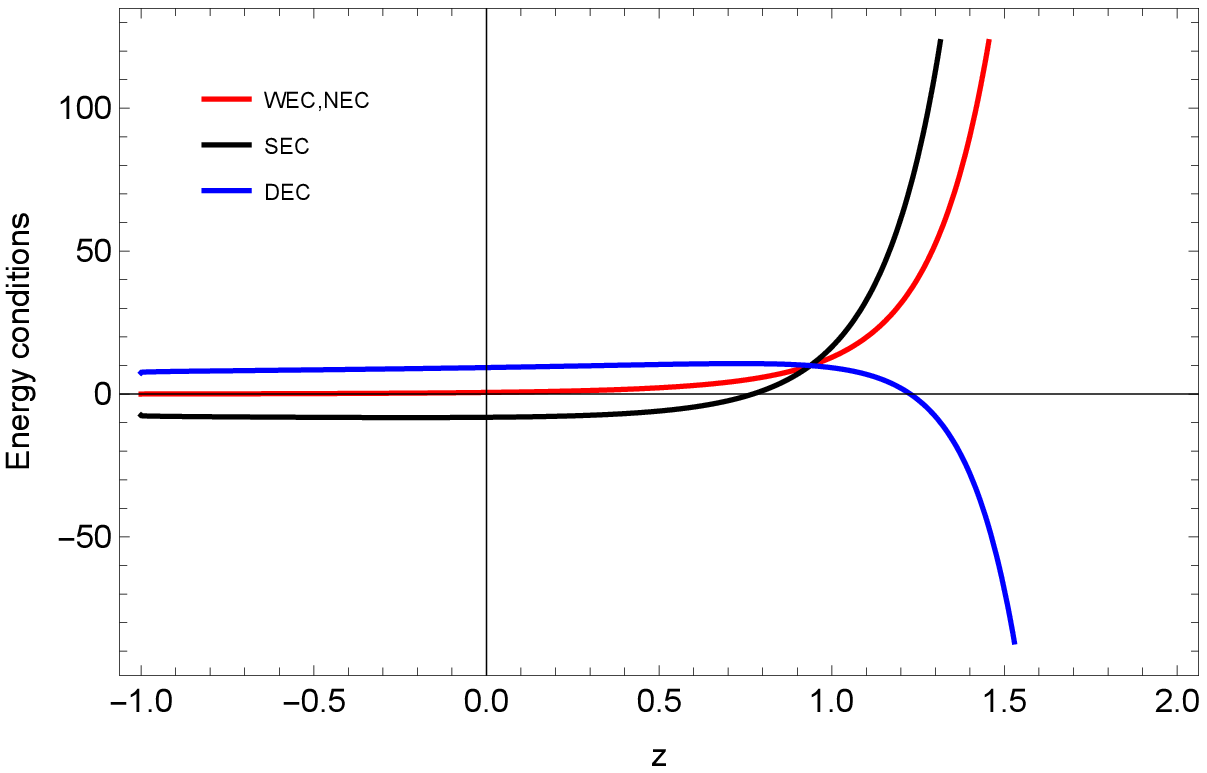}~~~~~~~~~
  \includegraphics[scale=0.5]{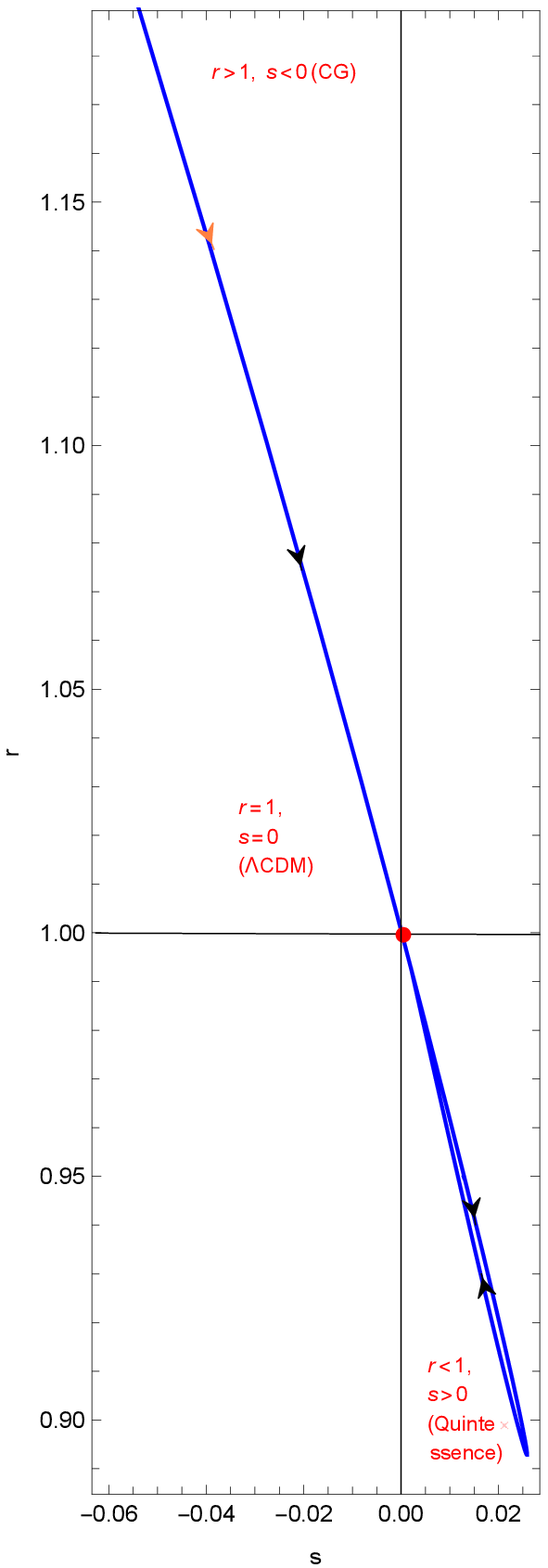}~~~~\\

{\bf{Figure. VIII:}} Energy conditions vs redshift $(z)$ for Model I (left plot) with $a_{1}=0.4$, $\delta=0.7$, $\sigma=1.5$, $\gamma=0.05$, $\alpha=1.2$, $\beta=1.5$ and $\lambda=0.005$, for Model II (right plot) with $a_{1}=0.4$, $\delta=0.7$, $\sigma=1.5$, $\gamma=0.05$, $\zeta=0.7$, $\psi=0.4$ and $\lambda=0.005$.
  \hspace{1cm}\vspace{5mm}

\vspace{3mm}
\vspace{3mm}

\end{center}
\end{figure}\\
For Model I, from equations (\ref{68}-\ref{69}),  we evaluated the expressions of WEC, NEC, DEC, SEC are as follows\\

WEC $\Rightarrow$
\begin{eqnarray}
 \rho_{de}+p_{de}=\phi_{0}^{2}\gamma^{2}a_{1}^{2\gamma}(t^{\delta}e^{\sigma t})^{2\gamma}\bigg(\frac{\delta}{t}+\sigma\bigg)^{2}+\frac{4\delta}{t^{2}}\bigg(\alpha+\frac{\beta}{12\bigg(\frac{\delta}{t}+\sigma\bigg)^{4}}\bigg) \geq 0,
  \rho_{de} \geq 0,
\end{eqnarray}
NEC $\Rightarrow$
\begin{eqnarray}\label{75}
 \rho_{de}+p_{de}=\phi_{0}^{2}\gamma^{2}a_{1}^{2\gamma}(t^{\delta}e^{\sigma t})^{2\gamma}\bigg(\frac{\delta}{t}+\sigma\bigg)^{2}+\frac{4\delta}{t^{2}}\bigg(\alpha+\frac{\beta}{12\bigg(\frac{\delta}{t}+\sigma\bigg)^{4}}\bigg) \geq 0,
\end{eqnarray}
DEC $\Rightarrow$
\begin{eqnarray}\label{76}
 \rho_{de}-p_{de}=2V_{0}e^{-\lambda\phi_{0}[a_{1}t^{\delta}e^{\sigma t}]^{\gamma}}+12\alpha\bigg(\frac{\delta}{t}+\sigma\bigg)^{2}-\frac{\beta}{2\bigg(\frac{\delta}{t}+\sigma\bigg)^{2}}-\frac{4\delta}{t^{2}}\bigg(\alpha+\frac{\beta}{12\bigg(\frac{\delta}{t}+\sigma\bigg)^{4}}\bigg) \geq 0,
\end{eqnarray}
SEC $\Rightarrow$
\begin{eqnarray}\label{77}\nonumber
 \rho_{de}+3p_{de} &=& 2\bigg[\phi_{0}^{2}\gamma^{2}a_{1}^{2\gamma}(t^{\delta}e^{\sigma t})^{2\gamma}\bigg(\frac{\delta}{t}+\sigma\bigg)^{2}\bigg]-2V_{0}e^{-\lambda\phi_{0}[a_{1}t^{\delta}e^{\sigma t}]^{\gamma}}-12\alpha\bigg(\frac{\delta}{t}+\sigma\bigg)^{2}\\
 &&+\frac{\beta}{2\bigg(\frac{\delta}{t}+\sigma\bigg)^{2}}+12\frac{\delta}{t^{2}}\bigg(\alpha+\frac{\beta}{12\bigg(\frac{\delta}{t}+\sigma\bigg)^{4}}\bigg) \geq 0.
\end{eqnarray}
The plot of energy conditions vs redshift is presented in Figure VIII. From Fig. VIII, we analyzed that WEC, NEC and DEC belongs to the positive region for all $z$ so they satisfied the conditions but SEC belongs to the negative region which shows that SEC is violated. Thus, the violation of SEC describes the accelerated expansion of the Universe. \\
\section{Statefinder Parameters}\label{sec6}
Several dark energy models have been made to describe the behavior of dark energy and the accelerating stage of the Universe. To characterize between these models, Sahni et. al., commenced a set of parameters which is called Statefinder parameter $\{r,s\}$ \cite{Sahni03}. The statefinder parameters $r$ and $s$ can be obtained as follows
\begin{equation}\label{78}
 r=\frac{\dddot a}{aH^{3}},
\end{equation}
\begin{equation}\label{79}
  s=\frac{(r-1)}{3(q-\frac{1}{2})}.
\end{equation}
The parameter $r$ in equation (\ref{78}) can be expressed as,
\begin{equation}\label{80}
r=2q^{2}+q-\frac{\dot{q}}{H}.
\end{equation}
 From equations (\ref{65}) and (\ref{66}), we evaluated the form of $r$ $\&$ $s$ are as follows,
 \begin{equation}\label{81}
   r=1-\frac{3\delta}{(\delta+\sigma t)^{2}}+\frac{2\delta}{(\delta+\sigma t)^{3}},
 \end{equation}
 \begin{equation}\label{82}
   s=\frac{\frac{2\delta}{(\delta+\sigma t)^{3}}-\frac{3\delta}{(\delta+\sigma t)^{2}}}{\frac{3\delta}{(\delta+\sigma t)^{2}}-\frac{9}{2}}.
 \end{equation}
 The different relations of the statefinder parameters $r$ $\&$ $s$ illustrate the several dark energy models which are as follows \cite{Alam03}\\
 $(1)$ $\Lambda$CDM model for $r=1$, $s=0$,\\
 $(2)$ Quintessence model for $r<1$, $s>0$,\\
 $(3)$ Chaplygin gas model for $r>1$, $s=<0$.\\
 \begin{figure}[h!]
\begin{center}
  \includegraphics[scale=0.6]{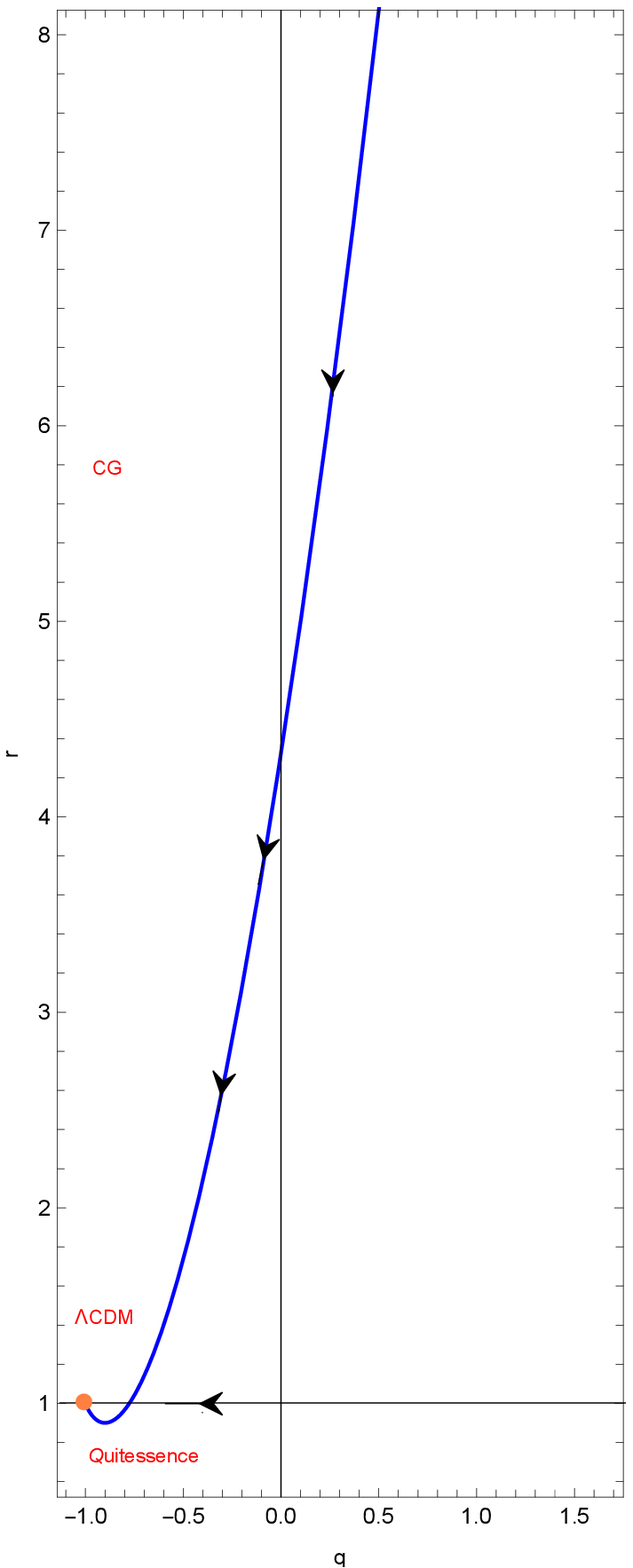}

{\bf{Figure. IX:}} Plot of r-s plane for $a_{1}=0.4$, $\delta=0.7$, $\sigma=1.5$ and $\gamma=0.05$.
  \hspace{1cm}\vspace{5mm}

\vspace{3mm}
\vspace{3mm}

\end{center}
\end{figure}\\
 \begin{figure}[h!]
\begin{center}
  \includegraphics[scale=0.6]{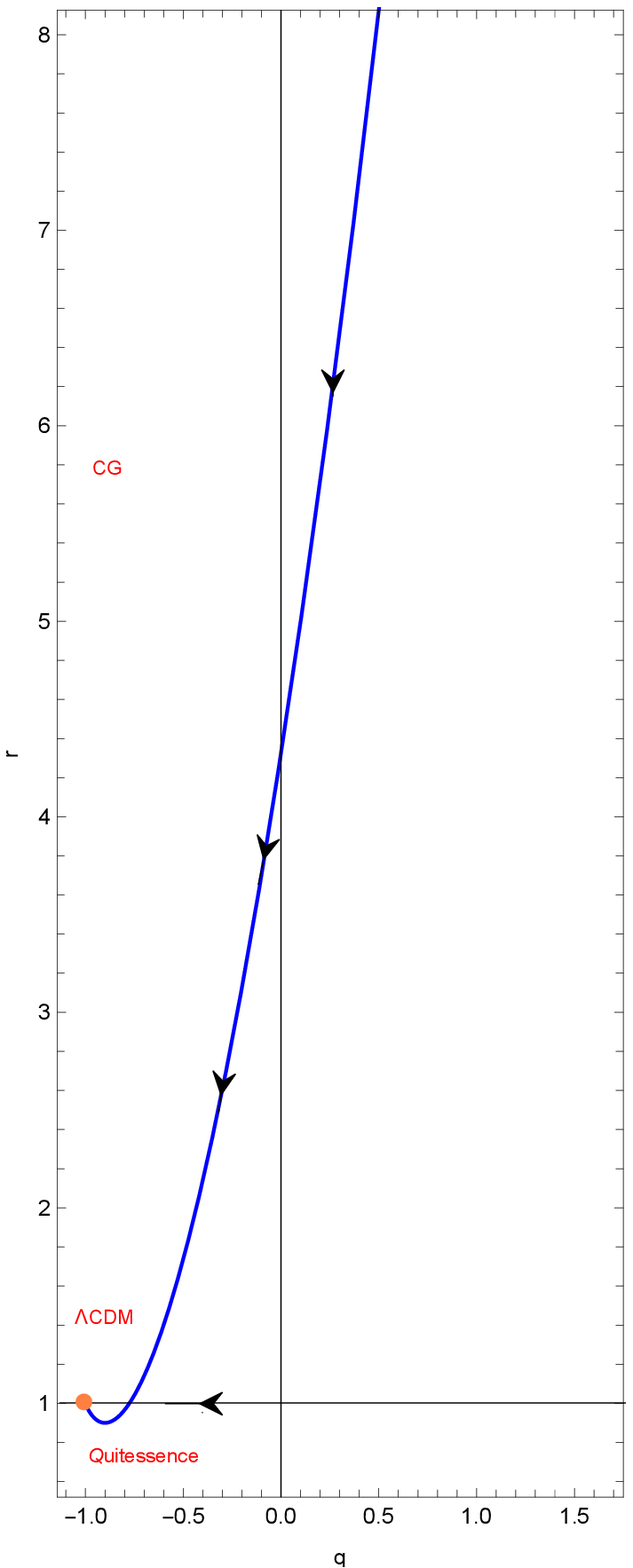}~~~~~~~~~

{\bf{Figure. X:}} Plot of r-q plane for $a_{1}=0.4$, $\delta=0.7$, $\sigma=1.5$ and $\gamma=0.05$.
  \hspace{1cm}\vspace{5mm}

\vspace{3mm}
\vspace{3mm}

\end{center}
\end{figure}\\
The plot of r-s for the parameters $a_{1}=0.4$, $\delta=0.7$, $\sigma=1.5$ and $\gamma=0.05$ are presented in Figure IX. From Figure IX, we observe that at initial stage, our model starts from the range $r>1$, $s<0$ which illustrate the Chaplygin gas model, after that our model represents quintessence model for $r<1$, $s>0$ and at the end, the model belongs to the point $\{r,s\}$= $\{1,0\}$. The characteristics of the statefinder parameters $\{r,s\}$ for the model be $r=1$, $s=0$ which illustrate the $\Lambda$CDM model. In Figure X, we plot the $r-q$ plane to discuss the behavior of our model for some suitable parameter. In this figure, the middle line shows the $\Lambda$CDM model and it divides the figure into two equal parts where the upper part of the line belongs to the Chaplygin gas model and the lower part of the line represents the quintessence model. The trajectories in $r-q$ plane starts at the point $q>0$ and $r>0$ which represents the SCDM model, after that it represents the quintessence model for $q<0$ and $r<1$ and at the end, it goes to the de-Sitter stage at $q=-1$, $r=1$.\\
\section{ Conclusions}\label{sec7}
We investigated a dynamical system analysis of $f(T,\phi)$ gravity theory. We assume two forms of $G(T)$  which are $(i)$ $G(T)$ = $\alpha T+\frac{\beta}{T}$, $(ii)$ $G(T)$ = $\zeta T$ ln$(\psi T)$ where $\alpha$, $\beta$, $\zeta$ and $\psi$ be the free parameters and analyze the stability behavior of these models by phase portrait. We evaluated the system of differential equations from Friedmann equations by introducing some new dimensionless variables $(x,y,z,r,\rho,\lambda,\sigma)$. To discuss the stability analysis, we found equilibrium points from the set of autonomous differential equations. We found eight equilibrium points for Model I which are $A\;(0,0,0,0,0)$, $B\;(0,0,0,1,0)$, $C\;(0,0,\nu,1-\nu,0)$, $D\;(\delta,0,0,1-\delta^{2},0)$, $E\;(\frac{\sqrt{\frac{3}{2}}}{\lambda},\frac{1}{\lambda}\sqrt{\frac{3}{2}},0,0,0)$, $F\;(0,\zeta,0,1-\zeta^{2},0)$, $G\;(0,\tau,\eta,1-\tau^{2}-\eta^{2},0)$, $H\;(\frac{\lambda}{\sqrt{6}},\sqrt{1-\frac{\lambda^{2}}{6}},0,0,0)$ and seven critical points for Model II are $A_{1}\;(0,0,0,\eta_{1},0)$, $B_{1}\;(0,0,\eta_{2},1-\eta_{2},0)$, $C_{1}\;(\eta_{3},0,0,1-\eta_{3}^{2},0)$, $D_{1}\;(0,0,0,1-\eta_{4}^{2},\eta_{4})$, $E_{1}\;(\frac{\sqrt{\frac{3}{2}}}{\lambda},\pm\frac{1}{\lambda}\sqrt{\frac{3}{2}},0,\eta_{5},0)$, $F_{1}\;(0,\eta_{6},0,1-\eta_{6}^{2},0)$ and $G_{1}\;(0,0,\eta_{7},\eta_{8},\pm\sqrt{3\eta_{7}+3\eta_{8}-3})$. The equilibrium points $B$ and $C$ are stable while $E$ is stable at $\lambda^{2}=3$, $F$ is stable at $\zeta^{2}=1$ and $G$ is stable at $\tau^{2}=2$, $\eta=-1$. For the points $B$, $C$, $F$ and $G$, $\Omega_{m}=\Omega_{r}=0$, $\Omega_{de}=1$ represent the dark energy dominant Universe and EoS parameters $\omega_{de}=\omega_{tot}=-1$, deceleration parameter $q=-1$ assure the accelerated phase of the Universe but for the point $E$, at $\lambda^{2}=3$, $\Omega_{m}=0$, $\Omega_{r}=0$, $\Omega_{de}=1$ represents the dark energy dominant Universe while deceleration parameter $q=\frac{1}{2}$, EoS parameters $\omega_{de}=\omega_{tot}=0$ represents the decelerated stage of the Universe. The equilibrium points $B_{1}$, $F_{1}$ and $G_{1}$ are stable points for the Model II. For these three points the density parameters $\Omega_{m}=\Omega_{r}=0$, $\Omega_{de}=1$ represent the dark energy dominant Universe, the deceleration parameter $q=-1$, EoS parameters $\omega_{de}=\omega_{tot}=-1$ assure the accelerated phase of the Universe. The current value of EoS parameter $\omega_{de}= -1.035^{+0.055}_{-0.059}$(Supernovae Cosmological Project), $\omega_{de}= -1.073^{+0.090}_{-0.089}$(WMAP+CMB), $\omega_{de}= -1.03\pm0.03$(Planck 2018) \cite{Aghanim20,Amanullah10,Hinshaw13} and deceleration parameter $q=-1.08\pm0.29$ \cite{Camarena20}. From Table 2 and Table 4, our obtained values of EoS parameter $(\omega_{de}$) and deceleration parameter $(q)$ of the above two $f(T,\phi)$ models satisfy the observational data.\\

We assume the scale factor as a function of hybrid expansion law. We rewrite all the physical parameters with redshift by using the equation $t(z)=\frac{\delta}{\sigma}W\bigg[\frac{\sigma}{\delta}\bigg(\frac{1}{a_{1}(1+z)}\bigg)^{\frac{1}{\delta}}\bigg]$. From several observational data, at present our Universe belongs to the accelerated stage. So, the deceleration parameter belongs to the range $-1\leq q\leq0$. For our model, we find the deceleration parameter by using the scale factor and the equation of time and redshift. We plot the deceleration parameter vs redshift in Figure IV for $a_{1}=0.4$, $\delta=0.7$ and $\sigma=1.5$. From Figure IV, the deceleration parameter takes the positive value at initial stage of the Universe and it takes the negative value at present and finally it goes to $-1$ at $z=-1$ this means at initial stage the Universe starts from transition phase and finally it goes to acceleration phase for $z=-1$ which satisfies the present observations. We represent the energy density vs redshift plot in Figure V for both the models by the suitable choice of parameters. In the beginning, the energy density takes positive value for all $z$ and it goes to zero at $z=-1$. Figure VI represents the nature of pressure vs redshift for both models. At initial stage, pressure takes the large number for $z$ and goes to zero at $z\rightarrow -1$. We get the value of energy density is positive for all $z$ and pressure is negative which describes the acceleration of the present Universe. The nature of EoS parameter vs redshift is given in Figure VII. The EoS parameter $\omega$ belongs to the quintessence phase at $z=0$ and it goes to $-1$ for $z=-1$. Also in Model I, we find the numerical value of EoS parameter is $\omega_{0}=-0.992$ for the parameters $a_{1}=0.4$, $\delta=0.7$, $\sigma=1.5$, $\gamma=0.05$, $\alpha=1.2$, $\beta=1.5$ and $\lambda=0.005$ and in Model II, we find the numerical value of EoS parameter is $\omega_{0}=-0.883$ which satisfy the current Planck observational data. We examined the behavior of the energy conditions vs redshift and the plot of energy conditions are given in Figure VIII. WEC, NEC and DEC are satisfied all the conditions but SEC is not satisfied the conditions. In Sec. 6, Figures IX, X represent the nature of statefinder parameters. In $r-s$ plane at initial stage our model belongs to the quintessence phase and at the end, it goes to the $\Lambda$CDM model. In $r-q$ plane our model belongs to the SCDM model and finally it goes to $\Lambda$CDM model. Thus, we can conclude that the present models satisfy all the observational data. \\

\end{document}